\documentclass[12pt]{article}
\pdfoutput=1
\usepackage{epsfig,amsfonts,amssymb}
\usepackage{hyperref}
\usepackage{cite}
\usepackage{cite}
\usepackage{graphicx}
\usepackage{epsfig}

\def\ZZZ{{\hbox{ Z\kern-1.6mm Z}}}
\def\RRR{{\hbox{ R\kern-2.4mm R}}}
\def\CCC{{\hbox{ C\kern-2.0mm C}}}
\def\zzz{{\hbox{z\kern-1mm z}}}

\newcommand{\vt}{\vartheta}

\newcommand{\qeq}{{\hbox{=\kern-2.3mm ? \kern.5mm }}}
\renewcommand{\qeq}{=}

\newcommand{\OO}{{\cal O}}

\newcommand{\wt}{\widetilde}

\newcommand{\NN}{{\cal N}}

\newcommand{\be}{\begin{equation}}
\newcommand{\ee}{\end{equation}}
\newcommand{\ben}{\begin{eqnarray}\displaystyle}
\newcommand{\een}{\end{eqnarray}}

\newcommand{\bea}[1]{\begin{eqnarray}\label{#1} }
\newcommand{\eea}{\end{eqnarray}}

\newcommand{\refb}[1]{(\ref{#1})}

\newcommand{\sectiono}[1]{\section{#1}\setcounter{equation}{0}}

\def\one{{\hbox{ 1\kern-.8mm l}}}
\def\zero{{\hbox{ 0\kern-1.5mm 0}}}

\newcommand{\kdot}{.}

\begin{document}

\baselineskip 24pt

\begin{center}
{\Large  \bf
Negative discriminant states in $\NN=4$ supersymmetric
string theories}

\end{center}

\vskip .6cm
\medskip

\vspace*{4.0ex}

\baselineskip=18pt

\centerline{\large \rm   Ashoke Sen }

\vspace*{4.0ex}

\centerline{\large \it Harish-Chandra Research Institute}
\centerline{\large \it  Chhatnag Road, Jhusi,
Allahabad 211019, India}

\vspace*{1.0ex}
\centerline{E-mail:  sen@mri.ernet.in, ashokesen1999@gmail.com}

\vspace*{5.0ex}

\centerline{\bf Abstract} \bigskip

Single centered BPS black hole solutions
exist only when the
charge carried by the black hole has positive discriminant.
On the other hand
the exact dyon spectrum in heterotic string theory
compactified on $T^6$ 
is known to contain states with negative
discriminant. We show that all of these negative discriminant
states can be accounted for as two
centered black holes.
Thus after the contribution to the index
from the two centered black
holes is subtracted from the total microscopic
index, the index for states with
negative discriminant  vanishes even for finite
values of charges, in agreement
with 
the results from the black hole side.
Bound state metamorphosis -- which
requires us to identify certain apparently different 
two centered configurations 
according to a
specific set of rules -- plays a crucial role in this
analysis. We also generalize these results to a
class of CHL string theories.

\vfill \eject

\baselineskip=18pt

\tableofcontents

\sectiono{Introduction and Summary} \label{sint}

We now have exact results for the appropriate 
supersymmetric
index carried by a class of dyons in a class of $\NN=4$
supersymmetric string 
theories in four dimensions\cite{9607026,0412287,0505094,0506249,
0508174,0510147,0602254,0603066,
0605210,0607155,0609109,
0612011,0802.0544,0802.1556,0803.2692}.\footnote{A
review of these results with all the sign factors corrected
can be found in \cite{0708.1270,1008.3801}.}
Furthermore the dependence of the index on the
asymptotic values of the moduli fields has also been
completely understood in these 
theories\cite{0702141,0702150,0705.3874,
0706.2363,0806.2337,0809.4258,0807.4451}.
These results have been used to test the correspondence
between the microscopic results and the 
macroscopic results based
on the analysis of quantum gravity in the near horizon 
geometry of the black hole, both at 
perturbative\cite{9607026,0412287,0605210,
0607155,0609109,1005.3044,0905.2686,1012.0265} and
non-perturbative\cite{0810.3472,0903.1477,0904.4253}
level. 
The analysis has also been extended
to compare the prediction for the sign of the 
index\cite{1008.4209}, 
weighted index\cite{0911.1563,1002.3857} 
etc. with complete success. In particular the black hole
prediction for the sign of the index was tested against the
microscopic prediction for finite values of the charges,
indicating that the macroscopic description holds beyond the
large charge limit.

\begin{table} {\small
\begin{center}\def\st{\vrule height 3ex width 0ex}
\begin{tabular}{|l|l|l|l|l|l|l|l|l|l|l|} \hline 
$(Q^2,P^2){\backslash} Q.P$  & 2 & 3 & 4 & 5 & 6 & 7
\st\\[1ex] \hline \hline
(2,2) &     648 & {\bf 327} & {\bf 0} & {\bf 0} & {\bf 0} 
& {\bf 0} 
\st\\[1ex] \hline
(2,4) &   { 50064} & {\bf 8376} & {\bf -648} & {\bf 0}
& {\bf 0} & {\bf 0}
\st\\[1ex] \hline
(2,6)  & { 1127472} & 130329 & {\bf -15600} 
& {\bf 972} & {\bf 0} & {\bf 0} \st\\[1ex] \hline
(4,4) &  { 3859456}
&  { 561576} & 12800 & {\bf 3272} & {\bf 0} 
& {\bf 0} \st\\[1ex] \hline
(4,6)    & 
{ 110910300}  &  
{ 18458000} 
&  { 1127472} & {\bf 85176} & {\bf -6404} & {\bf 0}
\st\\[1ex] \hline
(6,6)  & 
{ 4173501828}
&  { 920577636} & { 110910300} & { 8533821} & 
{ 153900}  & {\bf 26622}
\st\\[1ex] \hline
 \hline 
\end{tabular}
\caption{Some results for the index
$d(Q,P)$ in heterotic string
theory on $T^6$ in the chamber ${\bf R}$
for different values of $Q^2$, $P^2$ and
$Q . P$. The boldfaced entries are for charges for which the
discriminant is negative.
} \label{t1}
\end{center} }
\end{table}

In this paper we shall carry out yet another comparison of
the results from the microscopic and the macroscopic
sides at finite values of the charges. 
Let us for definiteness consider the
particular $\NN=4$ supersymmetric string theory obtained
by compactifying heterotic string theory on
$T^6$ and denote by
$(Q,P)$ the (electric, magnetic) charge vectors carried
by a state in this theory.
We can then define moduli independent T-duality invariant
inner products $Q^2$, $P^2$ and $Q\cdot P$. 
Let us further focus on charge vectors satisfying 
$I\equiv \gcd\{Q_i P_j - Q_j P_i\}=1$\cite{0702150}
for which $Q^2$, $P^2$
and $Q\cdot P$ are known to be the complete set of
T-duality invariants describing a charge 
vector\cite{0712.0043,0801.0149}.\footnote{The 
microscopic index has been computed from first
principles only for these states. For states with
$I>1$ the result for the index was guessed in 
\cite{0802.1556}
and  some justification for this result based on an effective
string picture of the states was proposed in 
\cite{0803.2692}. In principle our analysis of negative
discriminant states can also be extended to these states,
but in this paper we shall restrict our analysis to $I=1$
states.}
Single centered black
hole solutions\cite{9507090,9512031} 
are known to exist only when the discriminant 
$Q^2P^2-(Q\kdot P)^2$ is positive.
On the other hand when
we examine the microscopic result for the index, we 
often find non-zero results for the index for many charge
vectors with negative discriminant. Table \ref{t1}
shows examples of such negative discriminant states
(denoted by boldfaced entries)
computed in a specific chamber {\bf R}
in the moduli space defined in \S\ref{scount}.
Thus the question that arises is: what is the macroscopic
interpretation of these negative discriminant states?

\begin{table} {\small
\begin{center}\def\st{\vrule height 3ex width 0ex}
\begin{tabular}{|l|l|l|l|l|l|l|l|l|l|l|} \hline 
$(Q^2,P^2,Q.P)$  & constituents & index & total index \st\\[1ex] \hline \hline
(2,2,3) &     $(Q,Q) + (0, P-Q)$  & {324} & {327} \st\\
 &     $(2Q-P, 2Q-P) + ((P-Q), 2(P-Q))$ & {3} &  \st\\[1ex] \hline
(2,4,3) & $(3Q-2P, 3Q-2P) + (2(P-Q), 3(P-Q))$   &
24   & 8376 \st\\
& $(2Q-P, 2Q-P) + ((P-Q), 2(P-Q))$   &
576   &\st\\
& $(Q,Q) + (0, P-Q) $  &
7776  &\st\\ 
 [1ex] \hline
(2,4,4)  & $(Q,Q) + (0, P-Q)$  & -648 & -648 \st\\[1ex] \hline
(2,6,4) & $(2Q-P, 2Q-P) + ((P-Q), 2(P-Q))$   & -48
   & -15600\st\\
& $(Q,Q) + (0, P-Q) $  & -15552
  &\st\\ [1ex] \hline
(2,6,5)    & $(Q,Q) + (0, P-Q) $
 & 972 & 972
\st\\[1ex] \hline
(4,4,5 )  & $(2Q-P, 2Q-P) + ((P-Q), 2(P-Q))$   &  72
   & 3272  \st\\
& $(Q,Q) + (0, P-Q) $  &  3200
  & \st\\ [1ex] \hline
(4,6,5 )  & $(4Q-3P, 4Q-3P) + (3(P-Q), 4(P-Q))$   & 24
    & 85176   \st\\
    & $(3Q-2P, 3Q-2P) + (2(P-Q), 3(P-Q))$   & 576
    &    \st\\
& $(2Q-P, 2Q-P) + ((P-Q), 2(P-Q))$   & 7776
    &\st\\
& $(Q,Q) + (0, P-Q) $  & 76800
  &\st\\ 
 [1ex] \hline
(4,6,6 )  & $(2Q-P, 2Q-P) + ((P-Q), 2(P-Q))$   &  -4
   & -6404 \st\\
& $(Q,Q) + (0, P-Q) $  &  -6400
  &\st\\ [1ex] \hline
(6,6,7 )  & $(2Q-P, 2Q-P) + ((P-Q), 2(P-Q))$   &  972
   &  26622 \st\\
& $(Q,Q) + (0, P-Q) $  &   25650
  &\st\\ [1ex] \hline
 \hline 
\end{tabular}
\caption{This table shows, for various choices of
$(Q,P)$, the contribution to the index in the chamber ${\bf R}$
from two centered black holes. The entries in the last
column, giving the total index from all the two centered
configurations, agree with the entries in table \ref{t1}
for the corresponding values of $(Q^2,P^2,Q\kdot P)$,
showing that all the negative discriminant states in table
\ref{t1} can be accounted for as two centered
configurations. This is a general result that we shall
prove in this paper.} \label{t2}
\end{center} }
\end{table}

One point of view one might take is that for finite values of
the charges for which the results are given in table 
\ref{t1}, the description of the system as black holes
breaks down and so one should not try to compare the
results in table \ref{t1} with the black hole results.
However the point of view that we would like to 
advocate is
that the black hole description of the system continues to
hold even for finite charges although the correction to the
Bekenstein-Hawking formula due to $\alpha'$ and string loop
effects may become significant. 
In that case the absence of black holes with
negative discriminant is in
apparent conflict with the presence of the negative
discriminant states in the microscopic spectrum. 
One could still argue 
that the absence of negative discriminant 
black hole solutions
was derived in the classical supergravity theory,
and perhaps this is modified in
string theory. To explore this we shall use the description
of this system as a collection of wrapped D5, 
D3 and D1-branes
in type IIB string theory on $K3\times T^2$
and assume  
that the entropy of a supersymmetric black hole
is independent of the asymptotic values of the moduli.
In this case,
one can show, using
a simple scaling argument along the line of 
\cite{0908.3402}, that the classical Wald
entropy, if non-zero, will grow
quadratically when we scale all the charges together.
On the other hand the logarithm of the microscopic index
grows at most linearly with the charges. Thus having a
regular black hole solution after inclusion of $\alpha'$
corrections will be in conflict with the microscopic results
in the large charge limit.
One might still wonder if quantum corrections
could change the result, but as in
\cite{0908.3402} we shall call a configuration
a black hole only if it exists as a solution to the classical
equations of motion.
Thus the absence of black holes
with negative discriminant in classical string theory would
imply that the contribution to the index from single centered
black holes  vanishes even for finite charges,
and we must look for different macroscopic
configurations to
account for these states in the microscopic theory.

For a special class of charge vectors, carrying 
$Q^2=P^2=-2$
and arbitrary values of $Q\kdot P$, this question was
analyzed by Dabholkar, Gaiotto and Nampuri\cite{0702150},
where they found that in the domain in the moduli
space where these states exist, there are two centered
configurations -- with individual centers 
carrying charges $(Q,0)$ and $(0,P)$ --
having the same index. 
This allowed them to identify these 
negative discriminant states as these specific two centered
bound states, resolving the apparent contradiction arising
due to the absence of single centered black holes 
carrying
these charges. However the analysis of \cite{0702150} also 
raises some
questions:
\begin{enumerate}
\item Are there other two centered configurations carrying 
the same total charge  in the same chamber of the moduli
space? If so then they would also contribute to the index
and spoil the agreement. We shall indeed find that there
are apparently 
a series of other 2-centered configurations which
contribute to the index in the same chamber of the moduli
space. Some examples are bound states of 
centers carrying charges $(Q + (Q\kdot P) P, 0)$ and
$(-(Q\kdot P) P, P)$, bound states of centers
carrying charges $(0, P + (Q\kdot P) Q)$ and
$(Q, - (Q\kdot P) Q)$ etc.
\item The full spectrum computed in the microscopic
theory contains many other negative discriminant states
carrying different charge vectors.
Are all the negative discriminant states in the
microscopic spectrum accounted for
by the bound states of multi-centered black holes?
This was indeed the motivation of \cite{0702150} for studying
the special case described above.
\end{enumerate}

In this paper we shall analyze these questions in detail.
We shall find
that indeed all the negative discriminant states in the
microscopic spectrum
can be accounted for in the macroscopic description
precisely as due to two centered 
bound states.\footnote{Here the individual centers, being
half BPS states, are in the S-duality orbit of elementary
heterotic string states\cite{dh1}, and
could appear either as small black 
holes\cite{9504147,0409148}
or smooth 
solutions\cite{0011217,0012025,0109154,0202072,
0212210,0512053} 
depending on the duality frame
used for their description\cite{0908.3402}. 
See footnote \ref{fo1} for
an extended discussion on this.}
The contribution of the latter to the
index can be calculated exactly and matched with the
microscopic results.
When the centers carry
charge$^2\ge 0$ the
analysis is straightforward. However
we find that when one or both of
the centers carry charge$^2=-2$, as in the case of
\cite{0702150}, there is a
subtlety, -- two or more bound states which have 
apparently different constituents but the same total charge
and the same index,
must be identified according to a precise set of rules.
We call this bound state metamorphosis
-- a phenomenon
similar to but not the same as the one observed in
\cite{1008.3555}. 
An analogous phenomenon in the context of
$\NN=4$ supersymmetric gauge theories was 
discussed in  \cite{0712.3625}.
This identification is justified since for 
a center
with charge$^2=-2$ there is no genuine
black hole solution, -- the
metric produced by it agrees with that of a black hole
only far away from the core\cite{9911161}. 
Thus at finite separation
the identity of individual centers is expected to be lost.
We also show that in the special limit in which
some of the dyons may be interpreted as gauge theory
dyons, the prescription we use for identifying bound states
reproduces the known spectrum of quarter BPS dyons in
gauge theories, in agreement with the results found in
\cite{0712.3625}.

Our result thus
shows that a specific prediction of the black hole
description of quarter BPS states in $\NN=4$ supersymmetric
string theory -- absence of single centered black holes with
negative discriminant -- is borne out by the microscopic 
spectrum, not only in the large charge limit but also
for finite charges. This shows that the description of the
system as a black hole can give us useful
information far beyond the leading 
semi-classical approximation. In this context we would like
to note that in $\NN=2$ supersymmetric
theories the absence of negative discriminant states
in certain region of the moduli space can be argued
independently by showing that such states, if present, would
become massless at certain points in the interior of the
moduli space and hence would produce additional
singularities in the moduli space which are known to be
absent\cite{9807087}. 
However in $\NN=4$ supersymmetric string theories
negative discriminant states never become 
massless\footnote{This can be seen by examining the
BPS mass formula
$$ m_{BPS}^2 = {1\over \tau_2} \left[(Q_R - \tau_1 P_R)^2
+ \tau_2^2 P_R^2 + 2 \sqrt{Q_R^2 P_R^2 - 
(Q_R\kdot P_R)^2}\right] $$
where $\tau_1+i\tau_2$ is the axion-dilaton modulus and
the subscript $R$ denotes projection of the charge
vectors along the  six graviphoton directions. In order for
$m_{BPS}$ to vanish for finite non-zero $\tau_2$,
each of the three 
terms inside [~] must vanish.
Since the inner product matrix
between the the right handed
components of charges is positive definite,
this requires $Q_R$ and $P_R$ to vanish.
In that case $Q^2 = Q_R^2 - Q_L^2 = - Q_L^2$, and
similarly $P^2=-P_L^2$ and $Q\kdot P=-Q_L\kdot P_L$.
Thus the discriminant is given by 
$Q^2 P^2 - (Q\kdot P)^2=Q_L^2 P_L^2 -
(Q_L\kdot P_L)^2$, and this is manifestly non-negative.
This shows that the negative discriminant states in
$\NN=4$ supersymmetric string theories
cannot become massless in the interior of the
moduli space.} in the
interior of the moduli space and hence there does not
seem to be an
independent
argument for their absence.

The rest of the paper is organized as follows. In 
\S\ref{scount}
we review the known microscopic
results on the dyon spectrum in $\NN=4$ supersymmetric
string theories and their moduli dependence. In
\S\ref{sbound} we discuss the phenomenon of bound
state metamorphosis and our prescription for
identifying bound states with different constituents
when one or both the constituents carry charge$^2=-2$. 
In \S\ref{sneg} we
show that once we implement the rules of bound
state metamorphosis, 
all the negative discriminant states are
accounted for precisely as two centered bound states.
To illustrate this point we have displayed in table \ref{t2}
the specific origin of all the negative discriminant states
which arise in table \ref{t1}. As can be seen, the entries for
the total index given in the last column of table 
\ref{t2} match
precisely the results for the total index given in table
\ref{t1}.
In \S\ref{sgen} we discuss
generalization of our analysis to
CHL 
models\cite{9505054,9506048,9508144,9508154}.
We conclude with some general remarks in \S\ref{sconc}.

\sectiono{The dyon spectrum} \label{scount}

We consider a dyon carrying electric charge $Q$ and
magnetic charge $P$ in
heterotic string theory compactified on $T^6$. Here $Q$
and $P$ are each 28 dimensional vectors, normalized so that
the components $Q_i$ and $P_i$ are integers. Let 
\be \label{es1}
Q^2, \qquad P^2, \qquad Q\kdot P\, ,
\ee
be the SO(6,22) invariant inner products of these vectors.
Then, if 
\be \label{es2}
I\equiv \gcd\{ Q_i P_j - Q_j P_i, \quad 1\le i,j\le 28\}  =1\, ,
\ee
then the index $d(Q,P) \equiv Tr'_{(Q,P)}(-1)^F$,
computed using the microscopic description of the
system, is given by
\be \label{es3}
d(Q,P) = g\left({P^2\over 2}, {Q^2\over 2}, 
Q\kdot P\right)\, ,
\ee
where $g(m,n,p)$ are the Fourier expansion coefficients
of the inverse of the Igusa cusp form of 
weight 10\cite{igusa1,igusa2}:
\be \label{es4}
{1\over \Phi_{10}(\rho,\sigma, v)}
= \sum_{m,n,p} (-1)^{p+1}\, 
g(m,n,p) \, e^{2\pi i (m\rho + n\sigma
+ p v)}\, .
\ee
Here $Tr'_{(Q,P)}$ denotes trace over BPS states carrying
charges $(Q,P)$ and preserving four supersymmetries,
{\it after removing the trace over the fermion zero modes
associated with broken 
supersymmetries.}\footnote{Alternatively 
we can describe it as the sixth helicity
supertrace\cite{9611205, 9708062} 
appropriately normalized so that a single
supermultiplet with helicities ranging from $-1$ to $1$
gives a contribution of 1 to the index.}
$\Phi_{10}(\rho,\sigma, v)$ is a
well defined function in the
Siegel upper half plane:
\be \label{es4.5}
\rho_2, \sigma_2 > 0, \qquad \rho_2 \sigma_2- v_2^2>0,
\quad (\rho_2, \sigma_2, v_2) \equiv Im\, (\rho,\sigma, v)\, ,
\ee
and can be expressed as an
infinite product
\be \label{es5}
{1\over \Phi_{10}(\rho, \sigma, v)} 
=e^{-2\pi i(\rho+\sigma+v)}
\prod_{j,k,l\in\zzz, \, 
4kl-j^2\ge -1\atop
k,l\ge 0, j<0 \, \hbox{\tiny for}\, k=l=0 }
\left( 1 - e^{2\pi i (k\rho+l\sigma + j v)}\right)^{-c(4kl - j^2)}
\ee
where the coefficients $c(s)$ are defined via the equation
\be \label{es6}
8\, \sum_{i=2}^4 {\vt_i(\tau, z)^2 \over \vt_i(\tau, 0)^2}
= \sum_{n,j} c(4n - j^2) \, e^{2\pi i (n\tau + j z)} \, ,
\ee
$\vt_i$'s being the Jacobi theta functions.
It follows from \refb{es6} that
\be \label{es7}
c(s)=0  \quad \hbox{for} \quad s< -1\, , \qquad c(-1)=2\, .
\ee

It turns out that $g(m,n,p)$ defined through \refb{es4} are
ambiguous since $1/\Phi_{10}(\rho, \sigma, v)$ has poles
in the $(\rho,\sigma, v)$ plane, and hence the $(\rho,\sigma,v)$
space is divided into different domains with each domain
having its own Fourier expansion that converges in that domain.
We shall always work in the region where 
$\rho_2$, $\sigma_2$
and $|v_2|$ are large. In this region a term of the form
\be \label{es8}
\left( 1 - e^{2\pi i (k\rho+l\sigma + j v)}\right)^{-\alpha}
\, ,
\ee
has a convergent expansion in a power series in
$e^{2\pi i (k\rho+l\sigma + j v)}$ as long as
$(k\rho_2+l\sigma_2 + j v_2)>0$. On the other hand
if $(k\rho_2+l\sigma_2 + j v_2)<0$, then we must express
this as
\be \label{es9}
(-1)^{-\alpha} e^{-2\pi i \alpha(k\rho+l\sigma + j v)}
\left( 1 - e^{-2\pi i (k\rho+l\sigma + j v)}\right)^{-\alpha}\, ,
\ee
and expand this in a power series in 
$e^{-2\pi i (k\rho+l\sigma + j v)}$. These two different modes
of expansion will in general generate different Fourier
coefficients $g(m,n,p)$, leading to a dependence 
of $g(m,n,p)$ on the domain in $(\rho_2,\sigma_2,v_2)$
space where we carry out the expansion.

\begin{figure}
\begin{center}
\epsfysize 1.5in
\epsfbox{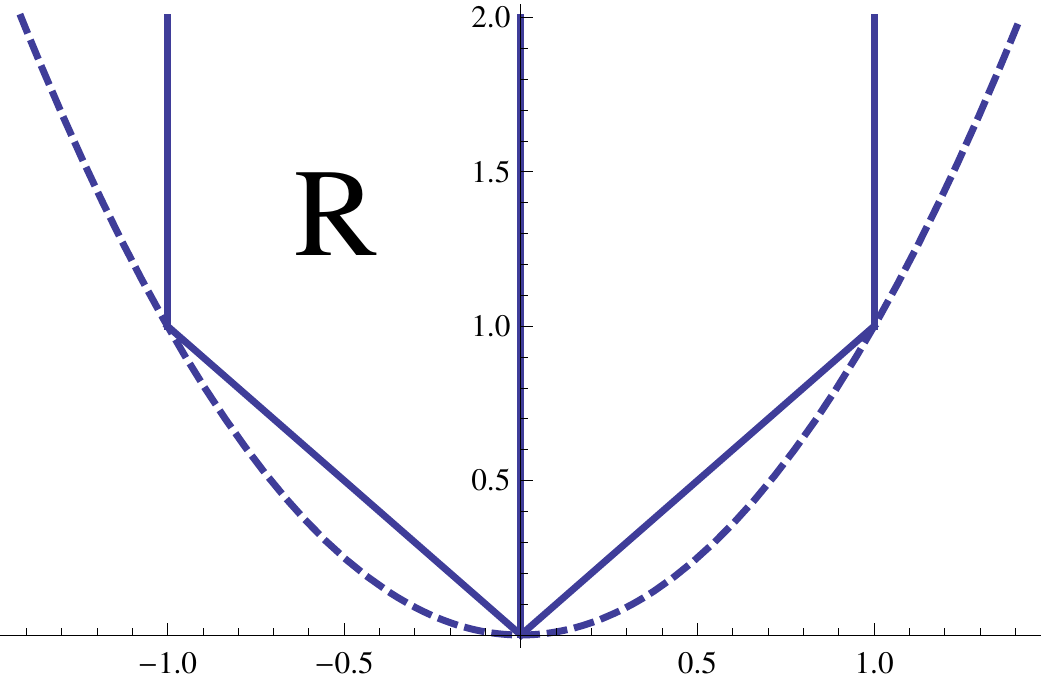}
\caption{The different domains 
in the $(\rho_2,\sigma_2,v_2)$ space. 
Here the $x$-axis labels
$(v_2/\sigma_2)$ and the $y$-axis labels
$(\rho_2/\sigma_2)$. The dashed line is the boundary
$\rho_2\sigma_2= v_2^2$ and the thick straight lines
label the walls across which $g(m,n,p)$ changes.
\label{f1}}
\end{center}
\end{figure}

Thus we need to examine for each $j,k,l$ in the product
in \refb{es5} whether $(k\rho_2+l\sigma_2 + j v_2)$ is
positive or negative and carry out the Fourier expansion
accordingly. It is straightforward to see that if 
$4kl - j^2\ge 0$, $k,l\ge 0$, 
then $(k\rho_2+l\sigma_2 + j v_2)$ is
always positive in the Siegel upper half plane described
in \refb{es4.5}, and hence there is no ambiguity in
expanding the corresponding term. Thus the only ambiguities
arise from the terms in the product for which $4kl - j^2=-1$.
For such terms the rules for the expansion changes
as we cross the plane
\be \label{es10}
k\rho_2+l\sigma_2 + j v_2 = 0, \quad j,k,l\in \ZZZ,
\quad  4kl-j^2 = -1\, ,
\ee
in the $(\rho_2, \sigma_2, v_2)$ space.
If we plot them in the $(v_2/\sigma_2, \rho_2/\sigma_2)$
plane, then \refb{es10} describes a set of straight lines
that divides the allowed region bounded by the parabola
$\rho_2/\sigma_2 > (v_2/\sigma_2)^2$ into infinite
number of traingles. Inside each triangle we shall have a
different set of $g(m,n,p)$. This has been illustrated in
Fig.~\ref{f1}.

\begin{figure}
\begin{center}
\epsfysize 1.5in
\epsfbox{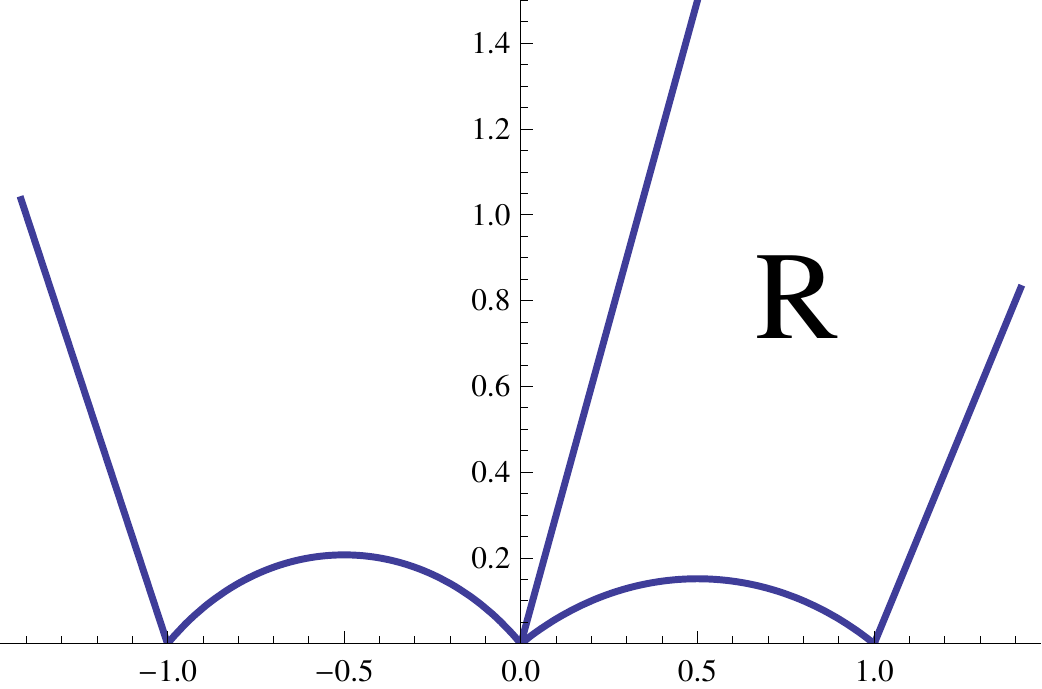}
\caption{The different domains 
in the complex $\tau$ plane separated by walls of
marginal stability. 
\label{f2}}
\end{center}
\end{figure}

It turns out that there is a one to one map between these
domains in the $(\rho_2,\sigma_2, v_2)$ space and the
chambers in the moduli space of heterotic string
theory on $T^6$, separated by the walls
of marginal 
stability\cite{0702141,0702150,0705.3874,0706.2363,
0806.2337}. 
To describe this we shall fix the
moduli associated with the metric, 2-form field and components
of gauge fields along $T^6$ -- labelled by points on the
coset space $SO(6,22)/SO(6)\times SO(22)$ -- to a fixed
value and study the walls in the upper half plane
parametrized by the axion-dilaton modulus $\tau$. The walls
of marginal stability turn out
to be  circles connecting rational points
$p/r$ and $q/s$, such that $p,q,r,s\in\ZZZ$ 
and $ps - qr=1$\cite{0702141}.
In the special case when $r$ (or $s$) vanishes, one of the
points is at infinity and the wall becomes a straight line
connecting an integer to $i\infty$. The precise shapes
of the walls depend on the charges $(Q,P)$.
As we cross a wall of marginal stability in the $\tau$ plane,
the index jumps and this is accounted for by a jump in
$g(m,n,p)$ across a wall of the form \refb{es10} in the
$(\rho_2,\sigma_2, v_2)$ plane, --
with each wall in the
$(\rho_2,\sigma_2, v_2)$ space being in one to one 
correspondence to a wall in the $\tau$ 
plane\cite{0702141}. In particular
the wall in the $\tau$ plane connecting the rational points
$p/r$ and $q/s$ is mapped to the wall
\be \label{ew1}
pq\sigma_2 + rs\rho_2 + (ps + qr) v_2 = 0
\ee
in the $(\rho_2,\sigma_2, v_2)$ space. This in particular
means that the chamber in the $\tau$ plane, bounded by
the walls connecting $(0, i\infty)$, $(0,1)$ and $(1,i\infty)$
gets mapped to the domain in the $(\rho_2,\sigma_2, v_2)$
space bounded by the walls $v_2=0$, $v_2=-\rho_2$
and $v_2=-\sigma_2$. We shall denote this chamber 
/ domain by ${\bf R}$. This has been marked in Figs.~\ref{f1}
and \ref{f2}.

$\Phi_{10}(\rho,\sigma,v)$ remains invariant under the
$SL(2,\ZZZ)$ S-duality group which
has a natural action on the
charges, $\tau$ as well as on $(\rho,\sigma, v)$ 
space.
It takes the form
\ben \label{ew2}
&& \pmatrix{Q\cr P}\to \pmatrix{a & b\cr c & d} 
\pmatrix{Q\cr P}, \qquad
\tau\to {a\tau + b\over c\tau + d}, \qquad
\pmatrix{a & b\cr c & d}\in SL(2,\ZZZ),  \cr\cr
&& \pmatrix{Q^2\cr P^2 \cr Q\kdot P}
\to \pmatrix{a^2 & b^2 & 2ab\cr c^2 & d^2 & 2cd\cr
ac & bd & (ad+bc)} \pmatrix{Q^2\cr P^2 \cr Q\kdot P},
\cr \cr
&&  \pmatrix{\sigma\cr \rho\cr v} \to
\pmatrix{d^2 & c^2 & - 2cd\cr b^2 & a^2 & - 2 ab
\cr -bd & -ac & (ad + bc)} \pmatrix{\sigma\cr\rho\cr  v}\, .
\een
This $SL(2,\ZZZ)$ transformation maps the walls in the
$\tau$ plane into each other and also the walls in the
$(\rho_2,\sigma_2, v_2)$ space into each other.
In fact just by knowing that the wall $v_2=0$ gets mapped
to the wall connecting $0$ and $i\infty$ in the $\tau$
plane we can derive \refb{ew1}, since
we can make use of the $SL(2,\ZZZ)$ transformation
generated by the matrix $\pmatrix{p & q\cr r & s}$ to 
map the $v_2=0$ wall to the wall \refb{ew1} in the
$(\rho_2,\sigma_2, v_2)$ space and the wall
connecting $0$ to $i\infty$ to the wall connecting 
$q/s$ to $p/r$ in the $\tau$ plane.

As we cross the walls of marginal 
stability in the $\tau$ plane shown in Fig.~\ref{f2},
certain bound states of two half BPS black holes 
either cease
to exist or come into existence thereby causing a jump in
the index\cite{0005049,0702146}.\footnote{Since 
each center represents a 
half BPS state, it may appear either as a small black
hole or as a smooth solution depending on the duality
frame in which we describe this\cite{0908.3402}. 
We shall call all of them
black holes. Since near a wall of marginal stability the 
distance between the centers go to infinity, a substructure
of the center will not affect the counting given in
\refb{ew3}. \label{fo1}}
Let us for definiteness consider the wall connecting
0 to $i\infty$ in the $\tau$ plane. If we take $Q\kdot P>0$, then
on the left of this wall we have a bound state of half BPS
black holes carrying charges $(Q, 0)$ and 
$(0,P)$\cite{0702150,0705.3874,0706.2363}, giving a
total contribution to the index 
\be \label{ew3}
(-1)^{Q\kdot P+1} \, |Q\kdot P| \, f(Q^2/2) \, f(P^2/2)\, ,
\ee
where $f(n)$ is defined through:
\be \label{ew4}
q^{-1} \prod_{k=1}^\infty (1-q^k)^{-24}
= \sum_{n=-1}^\infty f(n) q^n\, .
\ee
These two centered solutions cease to exist on the other
side of the wall, thereby causing a jump in the index
given by
$(-1)^{Q\kdot P} \, |Q\kdot P| \, f(Q^2/2) \, f(P^2/2)$
as we cross the wall from the left to the right. 
This can be shown to be equal to the jump in
$g(m,n,p)$ as we cross the corresponding wall in the
$(\rho_2,\sigma_2,v_2)$ space\cite{0702150,0705.3874,
0706.2363}.
For $Q\kdot P<0$
the situation is opposite, with the bound states existing on
the right of the wall. This can be seen by making an
S-duality transformation by $\pmatrix{0 & 1\cr -1 & 0}$
which
exchanges the two sides of the wall connecting 0 to
$i\infty$, and at the same time changes the 
sign of $Q\kdot P$.

By making an S-duality transformation of this result we
can also figure out the physical significance of the
other walls. In particular the S-duality transformation
by $\pmatrix{p & q\cr r & s}$ takes the
wall connecting
0 to $i\infty$ to the wall connecting $q/s$ to $p/r$
preserving orientation, \i.e.\ 0 gets mapped to $q/s$ and
$i\infty$ gets mapped to $p/r$.
Let us denote by $(Q',P')$ the transformed charges:
\be \label{ew6}
\pmatrix{Q'\cr P'} = \pmatrix{p & q\cr r & s} \pmatrix{Q\cr P}
\quad \Rightarrow \quad 
\pmatrix{Q\cr P} = \pmatrix{ s & -q \cr -r & p}
\pmatrix{Q'\cr P'}\, .
\ee
It then follows from S-duality
that if $Q.P>0$ \i.e.\ 
$(sQ'-q P')\kdot (-r Q' + p P')>0$
then to the left\footnote{Throughout this paper whenever
we refer to a wall connecting $A$ to $B$ we shall 
implicitly assign a orientation to the wall directed from 
$A$ to $B$, and the left or right of the wall is specified
with respect to this orientation.} of the wall
connecting $q/s$ to $p/r$
there exists a bound state of two centers carrying charges
\be \label{ew7}
\pmatrix{p & q\cr r & s}\pmatrix{Q\cr 0}
= \pmatrix{p(sQ'-q P')\cr r (sQ'-q P')} \quad
\hbox{and} \quad 
\pmatrix{p & q\cr r & s}\pmatrix{0\cr P}
= \pmatrix{q (-r Q' + p P')\cr s (-r Q' + p P')}\, .
\ee
On the other hand there are no such
bound states to the right of this wall. 
If $(sQ'-q P')\kdot (-r Q' + p P')<0$ then
the situation is reverse, with the bound state existing to
the right of the wall.

Now in \refb{ew7}  $(Q,P)$ and hence
$(Q',P')$ are arbitrary charge vectors.
Renaming $(Q',P')$ as $(Q,P)$ we can state the above
result
as follows.
Given a charge vector $(Q,P)$,
and a point in the $\tau$ plane, 
we have a bound state
of charges $(p(sQ-qP), r(sQ-q P))$ and 
$(q (-r Q+ p P), s (-r Q + p P))$ provided one of the following
two conditions hold:
\begin{enumerate}
\item $(sQ-qP).(-r Q+ p P)> 0$ and the point in 
the $\tau$ plane lies
to the left of the wall connecting $q/s$ to $p/r$.
\item $(sQ-qP).(-r Q+ p P) <0$ and the point in 
the $\tau$ plane lies
to the right of the wall connecting $q/s$ to $p/r$.
\end{enumerate}
In either case the net contribution to the index from this
bound state is given by
\be \label{ew5a}
(-1)^{Q\kdot P+1} \, |(sQ-qP).(-rQ+pP)| 
\, f((sQ-qP)^2/2) \, f((-rQ+pP)^2/2)\, ,
\ee
where in computing the
sign we have used the fact that $(-1)^{Q\kdot P}$ 
remains
invariant under S-duality transformation.
This allows us the determine the list of possible two
centered bound states
which can contribute to the index at a given point in the
moduli space and their contribution to the index. 
As we shall discuss in \S\ref{sbound} and
\S\ref{sneg}, this prescription
is modified somewhat when $(sQ-qP)^2$ and/or 
$(-rQ+pP)^2$ takes the
value $-2$.

\sectiono{Bound state metamorphosis} \label{sbound}

The prescription given at the end of \S\ref{scount} 
allows us to determine the total contribution to the index
from two centered bound states at any point in the
moduli space. We shall now argue that there are some
exceptions to this prescription when $(sQ-qP)^2$ or 
$(-rQ+pP)^2$ or both
are equal to $-2$ due to bound state metamorphosis,
-- a phenomenon first observed in \cite{0712.3625} in the
context of $\NN=4$ supersymmetric gauge theories.

\subsection{Prescription}

\begin{figure}
\begin{center}
\epsfysize=1.5in
\epsfbox{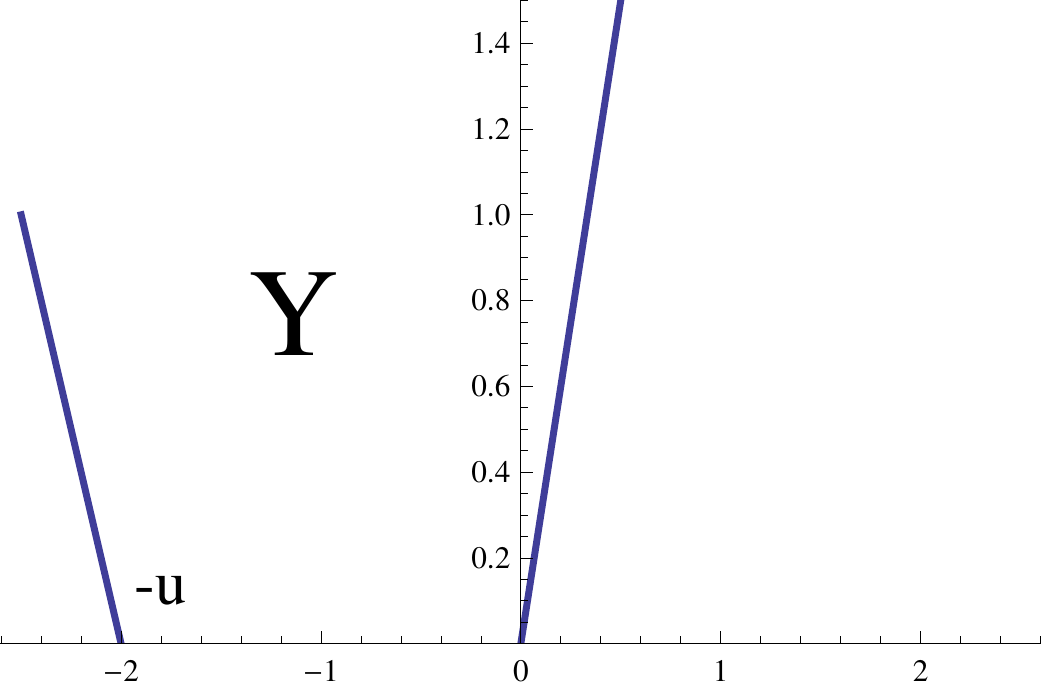}
\caption{For $P^2=-2$ and $u\equiv 
Q\kdot P>0$, the bound state
of $(Q,0)$ and $(0,P)$ exists to the left of the wall
connecting 0 and $i\infty$, while the bound state of
$(Q+u P, 0)$, $(-u P, P)$ exist to the
right of the wall connecting $-u$ and $i\infty$. We 
propose that
these two bound states are identical and exist in the
region Y between the two walls.
\label{f10}}
\end{center}
\end{figure}

We begin with the case when $P^2=-2$,
$Q.P>0$. In this case a two centered bound state of
$(Q,0)$ and $(0,P)$ exists to the left of the wall connecting
0 to $i\infty$, with a net contribution to the index given by
\refb{ew3}. Now consider the bound state of charges
$(Q+uP, 0)$ and $(-uP, P)$ for $u\equiv Q.P$. 
In the convention  described at 
 the end of \S\ref{scount}
this corresponds to the choice $\pmatrix{p & q\cr r & s}
=\pmatrix{1 & -u\cr 0 & 1}$ and
according to
the prescription given there, this
bound state exists to the right of the wall
connecting $-u$ to $i\infty$ (see Fig.~\ref{f10}) since
$(Q+uP)\kdot P=-Q\kdot P<0$. 
Furthermore \refb{ew5a} shows
that the contribution to the index from this bound state
is given by
\be \label{eab1}
(-1)^{Q\kdot P+1} |Q.P+uP^2| f((Q+uP)^2/2)
f(P^2/2)= (-1)^{Q\kdot P+1} |Q.P| f(Q^2/2)
f(P^2/2)\, ,
\ee
where we have used the
fact that $Q.P+uP^2 = - Q\kdot P$ and 
$(Q+uP)^2=Q^2$. Thus 
\refb{eab1} coincides with the \refb{ew3}, \i.e.\ 
the index carried by the bound state of $(Q,0)$ and $(0,P)$
and that carried by the bound state of $(Q+uP, 0)$
and $(-uP, P)$ coincide. We propose that these two
configurations describe the same physical states,
and hence should be counted only once. 
Furthermore
the bound state exists only in the region Y between
the walls connecting 0 and $i\infty$ and $-u$ and $i\infty$
(see Fig.~\ref{f10}). Had we not identified these bound states
then both two centered
bound states would exist in the region
between
the walls connecting 0 and $i\infty$ and $-u$ and $i\infty$
and one of the two centered 
bound states will exist in each of the two
regions to the right of the wall from 0 to $i\infty$ and to
the left of the wall from $-u$ to $i\infty$.

\begin{figure}
\begin{center}
\epsfysize=1.5in
\epsfbox{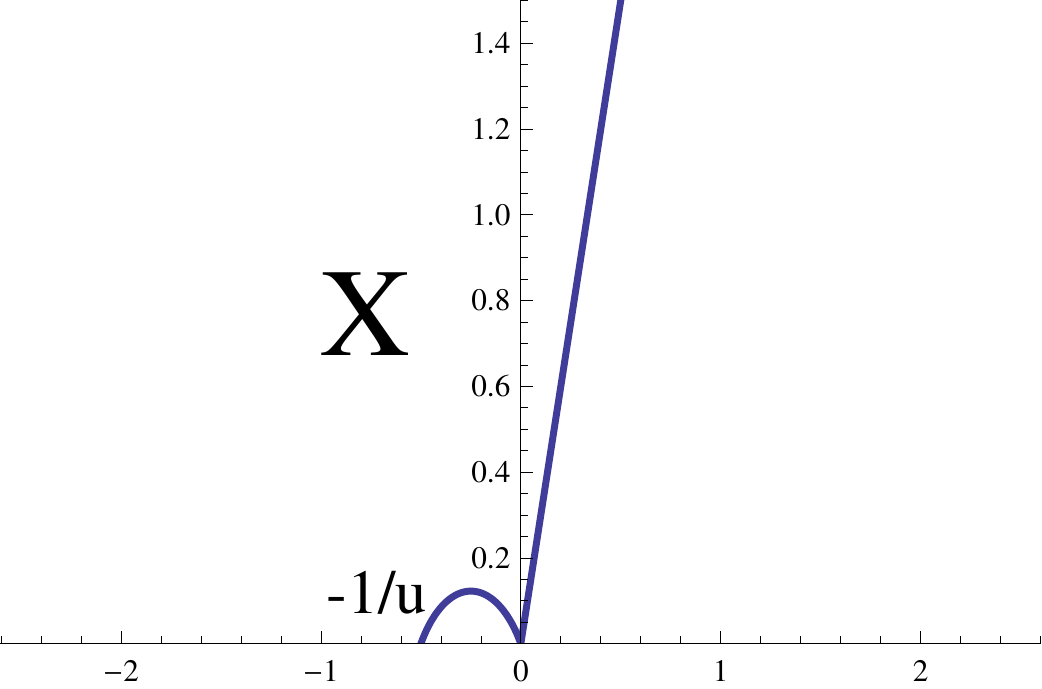}
\caption{For $Q^2=-2$ and $u\equiv
Q\kdot P>0$, the bound state
of $(Q,0)$ and $(0,P)$ exists to the left of the wall
connecting 0 and $i\infty$, while the bound state of
$(0,u Q+ P)$, $(Q,-u Q)$ exists to the
right of the wall connecting $0$ and $-1/u$. We propose that
these two bound states are identical and exists in the
region X between the two walls.
\label{f11}}
\end{center}
\end{figure}

If we have $Q\kdot P<0$ and $P^2=-2$ then the analysis
remains more or less unchanged with left and right
exchanged. Thus in this case there will be a bound state
of $(Q,0)$ and $(0,P)$ to the right of the wall connecting
0 to $i\infty$ and there will be a bound state of charges
$(Q+uP, 0)$ and $(-uP, P)$ to the left of the wall
connecting $-u$ to $i\infty$. Note however that now
$-u=-Q\kdot P$ is positive. Our prescription is to identify
these two bound states and postulate that it exists
in the region between the walls connecting $0$ to $i\infty$
and $-u$ to $i\infty$. Finally if $Q^2=-2$ and $P^2\ge 0$
then the whole
analysis can be repeated by exchanging $Q$ and $P$,
and we shall identify the bound state of $(Q,0)$ and $(0,P)$
with the bound state of $(Q, -uQ)$ and $(0, P+uQ)$,
existing in the region between the walls connecting
$0$ to $i\infty$ and 0 to $-1/u$  in the upper half 
$\tau$
plane (see Fig.~\ref{f11}).
In the convention described at the end of \S\ref{scount},
the second bound state corresponds to the choice
$\pmatrix{p & q\cr r & s} = \pmatrix{1 & 0\cr -u & 1}$.

\begin{figure}
\begin{center}
\epsfysize=1.8in
\epsfbox{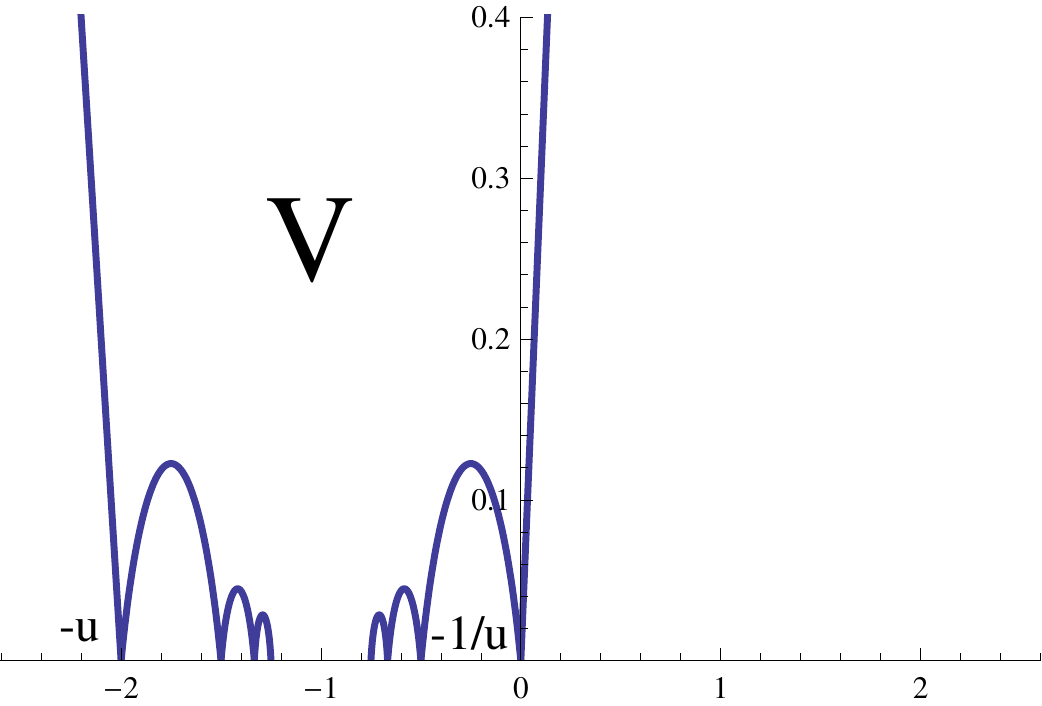}
\caption{For $Q^2=-2$ and $P^2=-2$, there are
a series of choices for  $(p,q,r,s)$
for which
the associated bound states all have the same index
and are identified. 
Thus these bound state
exist in a region V bounded by the walls 
corresponding to all
of these matrices.\label{f12}}
\end{center}
\end{figure}

If we have $P^2=Q^2=-2$ then we have to identify the
bound states of $(Q,0)$ and $(0,P)$, $(Q+uP, 0)$ and
$(-uP, P)$ and of $(0, uQ+P)$ and $(Q, -uQ)$, all of which
carry the same index. Furthermore it exists in the common
region lying between the walls connecting $0$ to $i\infty$,
$-u$ to $i\infty$ and $0$ to $-1/u$ (see Fig.~\ref{f12}).
However unless $u=\pm 1$
the identification does not stop here since
we can construct an infinite set of matrices 
$\pmatrix{p & q\cr r & s}$ by taking alternate products
of $\pmatrix{1 & -u\cr 0 & 1}$ and
$\pmatrix{1 & 0\cr u & 1}$ or of
$\pmatrix{1 & 0\cr -u & 1}$ and $\pmatrix{1 & u\cr 0 & 1}$,
and the bound states associated with all of these
matrices have the same index and should be identified.
Furthermore they exist in the region bounded by the walls
associated with these matrices, as shown by the region
V in 
Fig.~\ref{f12}.

Finally note that the analysis given above can be easily
extended to the case when neither $Q^2$ nor $P^2$ is
equal to $-2$, but for some given $SL(2,\ZZZ)$
matrix $\pmatrix{p & q\cr
r & s}$, either $(sQ- qP)^2=-2$ or $(-rQ + p P)^2=-2$ or
both are equal to $-2$. This is related to the cases
discussed above via a simple $SL(2,\ZZZ)$ transformation
by the matrix $\pmatrix{p & q\cr r & s}$, and
the region of the
moduli space where these bound states exist will be
related to the regions depicted in Figs.\ref{f10}, \ref{f11}
and \ref{f12} by this $SL(2,\ZZZ)$ transformation. 
We shall elaborate on this in \S\ref{sneg}.

\subsection{Justification}

Even though we have given a prescription for identifying
certain bound states which carry the same index, we have
not proved the result. 
We shall now try to provide some justification for this
prescription. This stems from the observation that 
a center with 
charge$^2=-2$ does not correspond to a conventional
black hole in any duality frame. 
In particular the central charge
associated to such a charge vector has a zero on a
subspace of the moduli space where the corresponding
state becomes massless, and the attractor flow 
cannot be continued past
this point to generate a horizon\cite{9807087,0005049}.
Beyond this point the solution is described by
a core in which the moduli are frozen at a constant
value at which the gauge symmetry is enhanced, 
and the charge carried by the black hole is spread
over the surface of this core\cite{9911161}. 
Thus the description of this state as a black hole is valid
only outside the core. Consequently a
two centered configuration where 
one or both the centers
carry charge$^2=-2$ can be regarded as a genuine two
centered solution only when the centers do not come too
close to each other.
Given that in the interior of the moduli space away from
the walls of marginal stability the centers do come
close to each other, it is not inconceivable that a two
centered solution of this type can change its description
as we move from one wall of marginal stability to another.
Since the index of a BPS state cannot jump
except at the walls of marginal stability, such a change
in the description is possible only if the new configuration
carries the same index as the old one. 

\begin{figure}
\begin{center}
\epsfysize=1.5in
\epsfbox{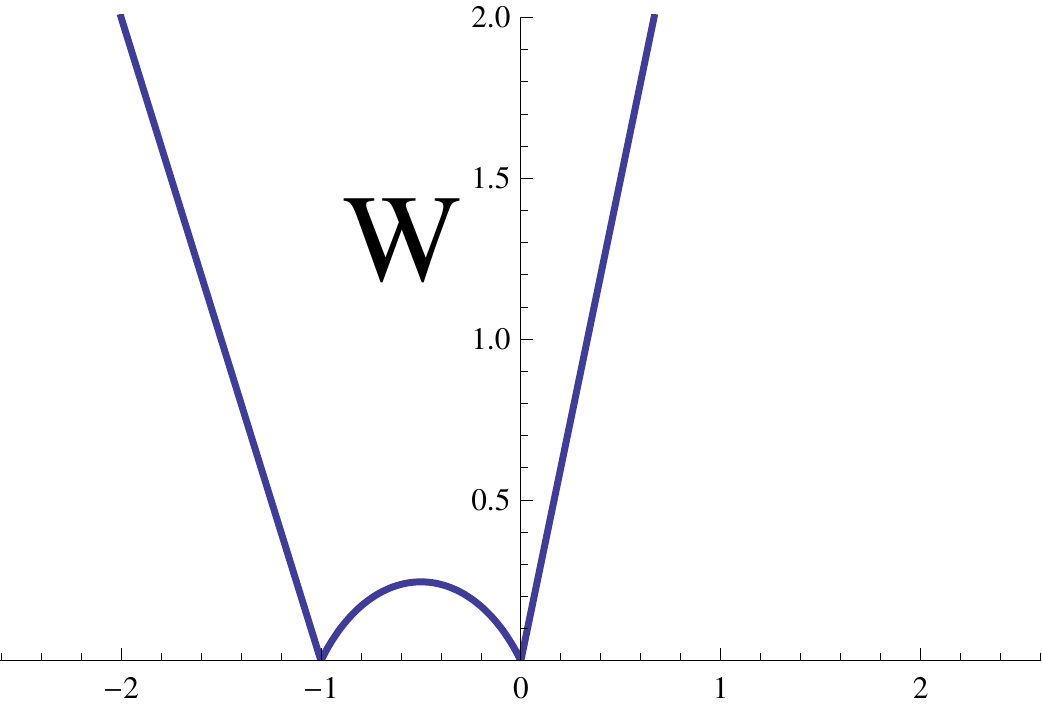}
\caption{For $Q^2=-2$, $P^2=-2$ and $Q.P=1$ 
the bound states of centers $(Q,0)$ and $(0,P)$,
$(Q+P, 0)$ and $(-P, P)$ and of
$(0, Q+P)$ and $(Q, -Q)$ all have the same index (=1)
and are identified. They exist in the region W
bounded by the
walls from 0 to $i\infty$, from $-1$ to $i\infty$ and from
0 to $-1$. 
This is consistent with the gauge theory limit in which
the BPS state carrying charge $(Q,P)$ with $Q^2=P^2=-2$
and $Q\kdot P=1$ is known to have index 1 and to exist
precisely in the chamber bounded by these three walls.
\label{f13}}
\end{center}
\end{figure}

While
the above argument provides a possible reason
for the metamorphosis, it clearly does not 
prove the validity of our prescription.
Presumably more insight into this can be obtained by
a detailed study of the two centered solution, but we shall
not do this here.
We shall see in \S\ref{sneg} that this
identification is necessary for a consistent description of the
negative discriminant states. 
In the remaining of this section we shall 
show that in the special case of $Q^2=P^2=-1$
and $Q\kdot P=\pm 1$ this identification is necessary
for getting agreement with the spectrum of quarter BPS
states in gauge theory. Let us choose $Q\kdot P=1$ for
definiteness. In this case our prescription
requires us to identify the bound states of
$(Q,0)$ and $(0,P)$, $(Q+P, 0)$ and $(- P, P)$,
and $(0, P +  Q)$
and $(Q, -  Q)$ in the
chamber W bounded by the walls connecting $0$ to $i\infty$,
$-1$ to $i\infty$ and  $0$ to $-1$ (see Fig.~\ref{f13}).
The index associated with each of these bound states
according to eq.\refb{ew5a} is 1. Thus if we follow
our prescription then the contribution
to the index from these bound states will be 
1 in the chamber $W$ in
Fig.~\ref{f13} and will vanish outside this
chamber. On the other hand if we do not make this
identification then the index will be 3 in the chamber $W$ and
2 outside this chamber (since only 2 of the 3 bound states
will exist in each of these regions). The correct
answer can be found by going 
near an appropriate region
of the moduli space where 
we can regard this state as a quarter
BPS state in an $\NN=4$ supersymmetric SU(3) gauge
theory\cite{0708.3715,0802.0761}. 
The state is known to have index 
1 and to exist only
inside the chamber bounded by the walls connecting 0 to
$i\infty$, $-1$ to $i\infty$ and 0 to 
$-1$\cite{9712211,9804160,9907090,0005275,0609055}. 
This will
be consistent with the result obtained from the spectrum
of two centered bound states only if we identify the 
different two centered bound states
as prescribed above. Thus at least the results in gauge theory
are in agreement with our general prescription.

\subsection{Uniqueness} \label{sunique}

In order to further justify that the bound state metamorphosis is
associated with centers carrying negative charge$^2$,
we shall now show that in order that two different bound states with
the same total charge have identical index and exist in the same region
of the moduli space, at least one of the centers must carry negative
charge$^2$.\footnote{I wish to thank the
referee for drawing my attention to this issue.} Let us consider two possible
bound states with total charge $(Q,P)$. Using S-duality we can
take one of them to be a
bound state of charges $(Q,0)$ and $(0,P)$; let the other have
centers carrying charges
$(p(sQ-qP), r(sQ-q P))$ and 
$(q (-r Q+ p P), s (-r Q + p P))$ for some $SL(2,\ZZZ)$ matrix
$\pmatrix{p & q\cr r & s}$. Using the freedom of changing the
sign of $p,q,r,s$ without changing the configuration, we shall choose
$p$ to be non-negative. These two bound states will 
manifestly carry the same
index if and only if~\footnote{We are not including the accidental
cases where the  factors in \refb{ew5a} are different for the two
bound states but the product is identical. Also we ignore
the other possibility where we equate $Q^2$ with $(-r Q + p P)^2$
and $P^2$ with $(sQ-qP)^2$ since this just corresponds to a redefinition
of $p,q,r,s$.}
\be \label{eex1}
Q^2 = (sQ-qP)^2, \quad P^2 = (-r Q + p P)^2, \quad Q.P
= \mp (sQ-qP).(-r Q + p P)\, .
\ee
Leaving out the trivial case where $Q.P=0$ and hence the
bound state index vanishes, and the cases $\pmatrix{p & q\cr r & s}
=\pmatrix{1 & 0\cr 0 & 1}$ or $\pmatrix{0 & 1\cr -1 & 0}$ for which
the two bound states are manifestly identical,
one can show as a consequence of \refb{eex1}
that 
\be \label{eex2}
p=s \qquad 2 \, p\, Q.P = r Q^2 + q P^2\, ,
\ee
or
\be \label{eex2a}
r\, Q^2 + q\, P^2 = 0, \qquad 2 \, q \, Q.P = (s-p) \, Q^2, \qquad
2\, r\, Q.P =
(p-s) P^2 \, .
\ee
The solutions given in \refb{eex2} and \refb{eex2a}
are related to the $-$ and $+$ signs
in \refb{eex1}. For \refb{eex2} if $Q.P$ is positive (negative) then
the first bound state exists to the left (right) of the wall connecting
0 to $i\infty$ and the second bound state
exists to the right (left) of the wall connecting $q/s$ to $p/r$.
On the other hand for \refb{eex2a}
if $Q.P$ is positive (negative) then
the first bound state exists to the left (right) of the wall connecting
0 to $i\infty$ and the second bound state
exists to the left (right) of the wall connecting $q/s$ to $p/r$.

First consider the case given in \refb{eex2}. If 
$p=s\ge 2$ then $qr=(ps-1)\ge 3$ and neither $q$ nor $r$ can vanish.
We now see from \refb{eex2} that for $Q^2$
and $P^2\ge 0$, at least one of $q/p=q/s$ and $r/p$ must be positive
(negative) for $Q.P$ positive (negative). Since $pqrs$ is positive,
$ps$ and $qr$ have the same sign and hence $q/s$ and $p/r$ also
have the same sign. Thus both $q/s$ and $p/r$ must
be positive (negative) for $Q.P$ positive (negative). 
Furthermore since $q$ and $r$ have the same
sign, $rs$ has the same sign as $q/s$ and hence ${p\over r}-{q\over s}={1
\over rs}$ has the same sign as $q/s$. Thus for $Q.P$ positive (negative)
we have ${p\over r}>{q\over s}$ (${p\over r}<{q\over s}$). Thus
we see that for $Q.P$ positive (negative), the first bound state
which lies to the left (right) of the wall connecting 0 and $i\infty$, and the
second bound state which lies to the right (left) of the wall connecting
$q/s$ to $p/r$, with $0<{q\over s}< {p\over r}$ (${p\over r} < {q\over s}<0$),
exist in non-overlapping regions of the moduli space. Thus there is no
sense in which we can identify them.

The special cases $p=s=1$ and $p=s=0$ can be treated easily. If
$p=s=0$ then we must have $q=-r=\pm 1$. In this case the second bound
state is identical to the first one with the centers exchanged and hence
we can ignore this case. For $p=s=1$ we must have $qr=0$.
For $q=0$ \refb{eex2} gives $2Q.P=rQ^2$. Thus if $Q^2>0$, then
 for $Q.P$ positive (negative),
we must have $r$ positive (negative)
and $p/r$ positive (negative). Thus the first bound state exists
to the left (right) of the line from 0 to $i\infty$ and the second
bound state exists to the right (left) of the line connecting 0 to $p/r$,
with no common region in which both bound states exist. The case $r=0$
gives similar result. Thus we cannot identify these different bound states.

Finally we turn to the case \refb{eex2a}. First consider the special
case $Q^2=P^2=0$. In this case if $Q.P\ne 0$ then we must have $q=0$ and
$r=0$ and hence $ps=1$, $p=s=1$. Thus the two bound states
are identical and we can skip this case. If $Q^2=0$, $P^2>0$
then we must have $q=0$ and hence again $ps=1$, $p=s=1$.
This in turn implies from \refb{eex2a} that $r$ must also vanish, and we again
have that the two bound states are manifestly identical. Thus we need to
consider the case $Q^2,P^2> 0$. In this case either $q$ and $r$ both vanish
with $p=s=1$
in which case we again have identical bound states, or $q$ and $r$ are both
non-vanishing and have
opposite sign. 
Restricting our attention to the latter case, we now
get from \refb{eex2a} that 
\be \label{eex3}
2Q.P = \left({s\over q}-{p\over q}\right) Q^2 = \left({p\over r} - 
{s\over r}\right)P^2\, .
\ee
Using $ps-qr=1$ and $qr\le -1$
we see that we must have $ps\le 0$. First
consider the case $qr\le -2$ so that $ps\le -1$, and $p,q,r,s$
are all non-zero.
\refb{eex3} shows that for $Q.P$ positive (negative) ${s\over q}-{p\over q}$ is
positive (negative) and also 
${p\over r} - 
{s\over r}$ is positive (negative). But since $ps<0$ we know that 
${s\over q}$ and
${p\over q}$ have opposite sign and ${s\over r}$ and
${p\over r}$ have opposite sign. Thus we must have
${s\over q}$ and hence ${q\over s}$ positive (negative), 
${p\over r}$  positive (negative) and
${s\over r}$ negative (positive). Furthermore
we have ${q\over s} - {p\over r} = -{1\over rs}$ positive (negative).
Thus the first bound state exists to the left (right) of the wall connecting
0 to $i\infty$ and the second bound state exists to the left (right)
of the wall connecting ${q\over s}$ to ${p\over r}$ with
$0<{p\over r} < {q\over s}$ (${q\over s} < {p\over r}< 0$). Thus there
is no common region where both bound states exist.

Finally we need to consider the special case $qr=-1$, $ps=0$. First suppose
$p=0$. In this case \refb{eex3} shows that for $Q.P$ positive (negative)
we must have ${s\over q}$ positive (negative). 
Thus the first bound state exists to the left (right) of the wall
connecting 0 and $i\infty$ and the second bound state exists to the left (right)
of the wall connecting ${q\over s}$ to 0, with $0< {q\over s}$
(${q\over s}< 0$). Thus again we see that there is no common region 
where both bound state exists. For $s=0$ we see from \refb{eex3}
that for $Q.P$ positive (negative) we have ${p\over r}$ positive
(negative). In this case the first bound state exists to the left (right) of the wall
connecting 0 and $i\infty$ and the second bound state exists to the left (right)
of the wall connecting $i\infty$ to ${p\over r}$, with $0< {p\over r}$
(${p\over r}< 0$). Again there is no common region 
where both bound state exists. 

This shows that as long as $Q^2\ge 0$ and $P^2\ge 0$, a bound state
of $(Q,0)$ and $(0,P)$ cannot have the same index as another bound state
with same total charge, with both the bound states existing in a common
region in the moduli space. Thus the phenomenon of bound state
metamorphosis takes place if and only if the at least one of the centers
carry negative charge$^2$. This is consistent with the fact that precisely
in this case the classical black hole solution becomes singular at a finite
distance away from the center, and the mechanism suggested in 
\cite{9911161} is necessary to modify the solution. Thus while a multi-centered
solution can be trusted for large separation between the centers, it cannot
be trusted everywhere in the moduli space. As we shall see in 
\S\ref{sneg}, this is also consistent with the microscopic results.

\sectiono{Negative discriminant states} \label{sneg}

Given a charge vector $(Q,P)$ we denote the discriminant
associated with the charge vector as
\be \label{ed1}
D(Q,P) = Q^2 P^2 - (Q\kdot P)^2\, .
\ee
The spectrum of heterotic string theory on $T^6$, given by
eqs.\refb{es3}, \refb{es4}, contains many states with
negative discriminant. These arise from non-zero values of
$g(m,n,p)$ for $(m,n,p)$ satisfying $4mn-p^2<0$, and in turn
arise from the fact that the product over $k,l,j$ in \refb{es5}
contains terms with $4kl - j^2 <0$. 
Explicit examples of such negative discriminant states
can be found in the boldfaced entries in table \ref{t1}
where we have given the values of the index
$d(Q,P)$ for various
combinations of $Q^2$, $P^2$ and $Q\kdot P$ in the
chamber ${\bf R}$ depicted in Figs.~\ref{f1} and \ref{f2}.
On the other hand single centered black hole 
solutions\cite{9507090,9512031} always
have positive discriminant. Our goal in this section will
be to argue that all the negative discriminant states
in heterotic string theory on $T^6$ arise from bound states
of two half BPS black holes. Thus after subtracting the
contribution from the two centered configurations, we 
shall be
left with only states with non-negative discriminant, in 
agreement with the prediction based on classical black
hole solutions.

\subsection{Duality transformation}

In order to prove this result we shall use
the fact that the spectrum is consistent with wall
crossing, so that the jump in the spectrum as we cross a
wall of marginal stability can be explained as due to 
(dis-)appearance of a particular 2-centered solution. Thus if
we can establish the result in any one chamber, then
this will prove the result in all other chambers. 
We shall also 
make use of the duality
invariance of the spectrum, which tells us that if we can prove
this for any charge vector, then it also holds for all other
charge vectors related to the original charge vector
by a duality transformation.

Let us begin by making use of the duality transformation
to bring a charge vector 
$(Q,P)$ with $Q^2 P^2 < (Q\kdot P)^2$ to a
`standard form'. 
This is done using
the following steps:
\begin{enumerate}
\item
If either $Q^2$ or $P^2$ is $\le 0$ 
we skip this step and proceed directly to step 2.
Suppose however that both $P^2$ and $Q^2$
are positive. Let us take
for definiteness $P^2 \ge Q^2>0$. In this case
$|Q\kdot P|$ must be bigger than $Q^2$. Now let us
make a duality transformation $P\to P+ a Q$, $Q\to Q$
for some integer $a$. Under this $Q^2\to Q^2$ and
$Q\kdot P\to Q\kdot P
+ a Q^2$. Thus by choosing a suitable $a$ we can 
ensure that $|Q\kdot P| \le Q^2$. Under this duality transformation the new $P^2$ must be less than 
$Q^2$ and $|Q\kdot P|$ since otherwise
the discriminant will not be negative. If $P^2\le 0$ we proceed
to step 2. Otherwise we repeat the
process with $Q$ and $P$ exchanged so that $|Q\kdot P|$
becomes less than $P^2$. By continuing this process we can
ensure that either $Q^2$ or $P^2$ eventually
becomes less than or
equal to zero. 
\item At the end of step 1, we shall have a charge vector
for which either $Q^2\le 0$ or $P^2\le 0$. Using 
$Q\to P$, $P\to -Q$ symmetry we can ensure 
that $Q^2\le 0$. We now make a
transformation $P\to P+b \, Q$ so that
$P^2 \to P^2 + b^2 Q^2 + 2 \, b \, Q\kdot P$. 
If $Q^2< 0$,
then by  choosing $b$
to be sufficiently large we can make $P^2$ arbitrarily large
and negative. If on the other hand $Q^2=0$ then by
choosing $b$ to be sufficiently large in magnitude, and
having a sign opposite to that of $Q\kdot P$ we can
again make $P^2$ sufficiently large and negative.
We shall use this freedom to choose $P^2\le -4$ and
call this the standard form.
\end{enumerate}

Next we focus on the choice of the chamber in which we shall
work. We shall work in the chamber ${\bf R}$ shown in
Figs.~\ref{f1} and \ref{f2}. In the $(\rho_2, \sigma_2, v_2)$
plane this corresponds to choosing $-\sigma_2, -\rho_2
< v_2 < 0$. In this case  one can show that
\be \label{eb1}
k\rho_2 + l\sigma_2 + j v_2 > 0 
\ee
for the range of values of $j,k,l$ over which the product
runs in \refb{es5}. Thus in each factor of the product
we need to expand the $\left(1 - 
e^{2\pi i(k\rho+l\sigma+jv)}
\right)^{-c(4kl-j^2)}$ factor in $1/\Phi_{10}$ in powers of
$e^{2\pi i(k\rho+l\sigma+jv)}$ with $k,l\ge 0$ and
$j<0$ if $k=l=0$.
This in particular will mean that $g(m,n,p)$ vanishes for
$m< -1$ or $n<-1$. Since the standard form in which we
have brought $(Q,P)$, we have $P^2\le -4$, it follows that 
$d(Q,P)=g(P^2/2, Q^2/2, Q\kdot P)$
vanishes in the chamber ${\bf R}$. Thus if we can prove that
in this chamber there are no two centered configurations
contributing to the index, then we would have proved that
in any chamber the index for negative discriminant states
vanish after removing the contribution from the two
centered configurations.

\begin{figure}
\leavevmode
\begin{center}
\hbox{
\epsfysize=1.5in
\epsfbox{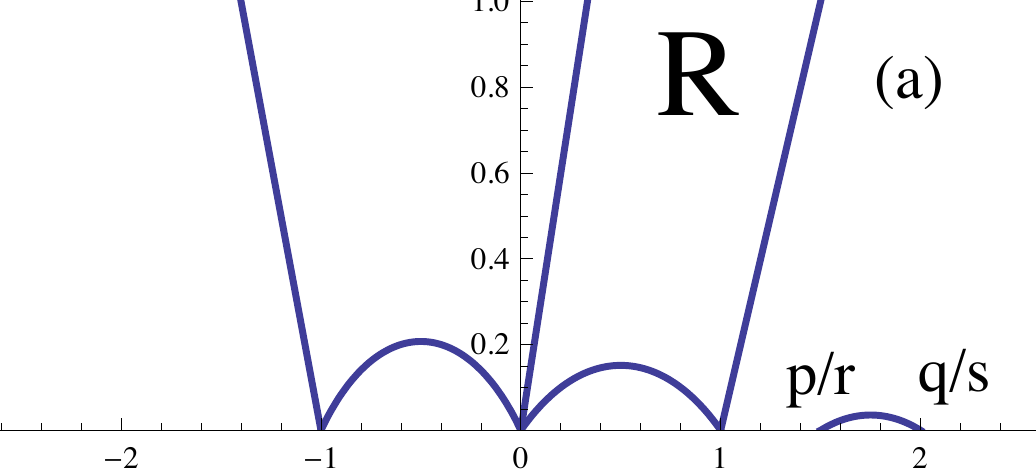} 
\epsfysize=1.5in
\epsfbox{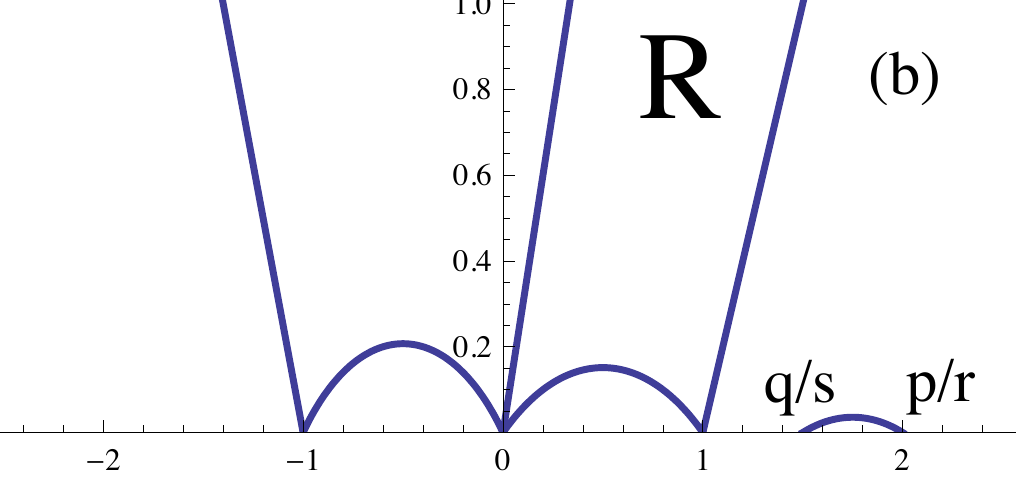}
}
\end{center}
\caption{This figure displays two possibilities for
how the chamber ${\bf R}$ is situated with respect to the wall
connecting $q/s$ to $p/r$ in the $\tau$ plane.}
\label{f3}
\end{figure}

Thus our task now is to analyze the two centered bound states
which could contribute to the index $d(Q,P)$ in the chamber
${\bf R}$. As discussed
at the end of \S\ref{scount}, the possible charges carried
by the two centers are of the form
\be \label{eb2}
(p \wt Q, r\wt Q), \quad (q \wt P, s \wt P)\, ,
\quad \wt Q = sQ - qP, \quad \wt P = -rQ + p P, 
\quad \pmatrix{p & q\cr r & s} \in SL(2,\ZZZ),
\quad s\ge 0\, ,
\ee
where the last condition is chosen using the freedom
of changing the sign of $\wt Q$, $\wt P$ and
$(p,q,r,s)$ simultaneously.
It follows
from \refb{eb2} that
\be \label{eb3}
P^2 = r^2 \wt Q^2 + s^2 \wt P^2 + 2 rs \wt Q\kdot \wt P\, .
\ee
The net index carried by such a bound state is given
by
\be \label{et1}
(-1)^{\wt Q\kdot \wt P+1} |\wt Q\kdot \wt P|\,
f(\wt Q^2) f(\wt P^2)\, .
\ee
It was also shown at the end of \S\ref{scount} that in the
$\tau$ plane these
bound states exist to the left of the wall connecting 
$q/s$ to $p/r$ if $\wt Q\kdot \wt P>0$ and to the right
of the same wall if $\wt Q\kdot \wt P<0$. 
We shall now analyze various possibilities separately.

\subsection{$\wt Q^2\ge 0$, $\wt P^2 \ge 0$}
Since $P^2<0$, it follows from \refb{eb3} that 
we must have $rs<0$ for $\wt Q\kdot \wt P >0$
and $rs> 0$ for $\wt Q\kdot \wt P<0$. Now since $ps-qr=1$
we have $ps > qr$. Dividing both sides by $rs$ we see that
for $\wt Q\kdot \wt P >0$ we have $p/r < q/s$ and hence
in the $\tau$ plane the chamber ${\bf R}$ lies to the right of the 
wall connecting $q/s$ to $p/r$ (see Fig.\ref{f3}(a)), 
-- precisely opposite to
the side in which bound states exist. On the other hand
for $\wt Q\kdot \wt P<0$ we have $rs>0$, $p/r> q/s$, and
hence the chamber ${\bf R}$ lies to the left of the wall connecting
$q/s$ to $p/r$ (see Fig.\ref{f3}(b)), 
-- again on the side opposite to which the bound
state exists. 
For $\wt Q\kdot \wt P=0$ there is no contribution to
the index from such a bound state (see eq.\refb{et1}), so
we need not consider this case.
This shows that in the chamber ${\bf R}$ there are
no bound states of charges of the form given in
\refb{eb2} as long as $\wt Q^2\ge 0$ and $\wt P^2\ge 0$.

\begin{figure}
\begin{center}
\epsfysize=2in
\epsfbox{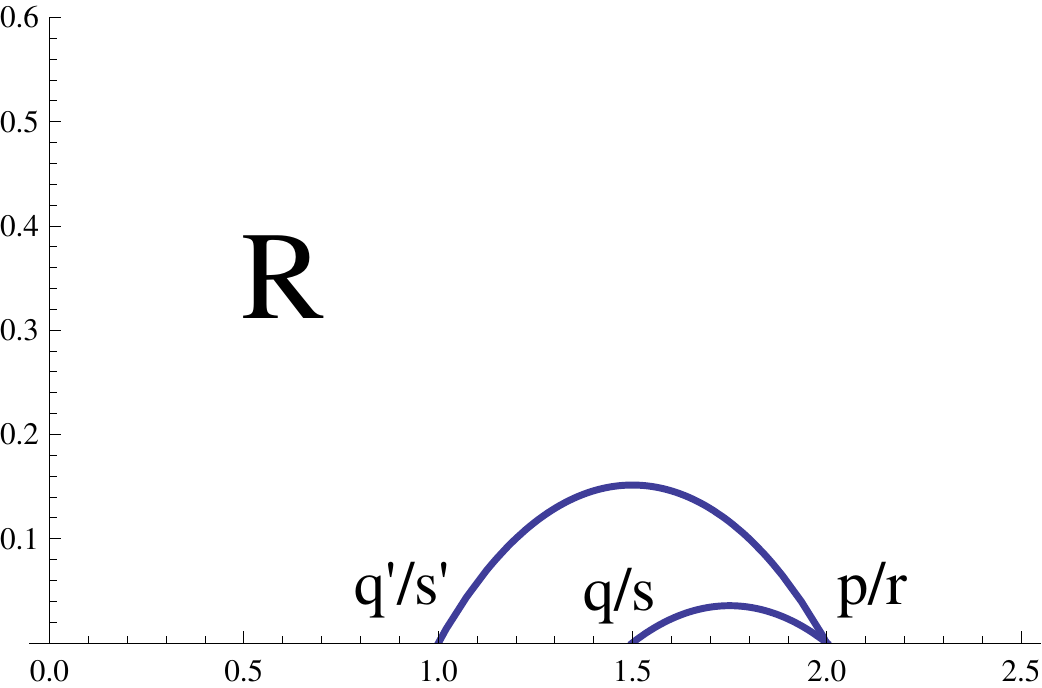}
\caption{The two walls of marginal stability corresponding
to bound states of charges given in eqs.\refb{eb2} and
\refb{eb22} for $\wt P^2=-2$, $\wt Q\kdot \wt P>0$. 
The bound states exists in the region between the two walls,
and hence not in ${\bf R}$.
\label{f4}}
\end{center}
\end{figure}

\subsection{$\wt P^2=-2$, $\wt Q^2\ge 0$ or $\wt Q^2=-2$, 
$\wt P^2\ge 0$}
Consider first the case when 
$\wt P^2=-2$, $\wt Q^2 \ge 0$, $\wt Q\kdot \wt P>0$. In
this case if $rs<0$ then the argument of the previous
paragraph establishes that there is no bound state
of charges given in \refb{eb2} in the chamber ${\bf R}$. Thus
we need to analyze the case when $rs\ge 0$. 
First consider the case $rs>0$ so that we have
$p/r > q/s$. 
In this case the chamber ${\bf R}$ lies to the left
of the wall connecting $q/s$ to $p/r$, -- the same side
on which we have a two centered solution carrying the
charges given in \refb{eb2} (see Fig.~\ref{f4}). 
This would seem to indicate that in the chamber ${\bf R}$
there is a bound state of the charges given in
\refb{eb2} carrying total index
given in \refb{et1}.
We shall however argue that we encounter another wall 
before reaching the chamber ${\bf R}$ across which this
bound state ceases to exist. 
To see this let us
examine another bound state carrying total charge
$(Q,P)$ where the individual centers carry charges
\ben \label{eb22}
&& (p' \wt Q', r'\wt Q'), \quad (q' \wt P', s' \wt P')\, ,
\qquad \wt Q' = s'Q - q'P, \quad \wt P' = -r'Q + p' P, 
\nonumber \\ \cr
&& \pmatrix{p' & q'\cr r' & s'} =  \pmatrix{p & q-pu\cr
r & s - ru}, \quad \pmatrix{\wt Q'\cr \wt P'} = \pmatrix{\wt Q+  u
\wt P\cr \wt P}, \quad u\equiv \wt Q\kdot \wt P.
\een
This has the same index as \refb{et1}, and
according to the bound state metamorphosis proposal
of \S\ref{sbound} this bound state should be identified
with the one given in \refb{eb2}.
Now 
it follows from \refb{eb3} and the $P^2<0$ 
condition that
\be \label{et2}
s / r > \wt Q\kdot \wt P\, .
\ee
{}From \refb{et2}, \refb{eb22} we get
\be\label{et3}
{q'\over s'} = {q - pu\over s - ru} 
< {q\over s}\, ,
\ee
as shown in Fig.~\ref{f4}.
Thus the wall connecting $q'/s'$ to $p'/r'=p/r$ lies in
between the wall connecting $q/s$ to $p/r$ and the chamber
${\bf R}$. Furthermore since
$\wt Q'\kdot \wt P'=-\wt Q\kdot \wt P<0$, the bound state
of the charges given in \refb{eb22} exists to the right 
of the wall connecting $q'/s'$ to $p/r$. 
Thus according to the prescription of
\S\ref{sbound}
such a
bound state will exist only in the region bounded
by the walls connecting $q/s$ to $p/r$ and $q'/s'$ to $p/r$
(see Fig.~\ref{f4}). In particular it will not exist in
the chamber ${\bf R}$.

Next consider the case where $rs=0$. It follows  from 
\refb{eb3} and that $\wt Q^2\ge 0$, $P^2\le -4$ that
$s$ cannot vanish, hence $r$ must vanish. In this case
we have $ps=1$ and hence $p=s=\pm 1$. Eq.\refb{eb3}
now gives $P^2=-2$ which contradicts the fact that $P^2$
has been chosen to be $\le -4$. Thus we see that $rs$
cannot vanish.

The case when $\wt Q\kdot \wt P<0$ can be dealt with in the
same way as in the previous case with some changes in
sign. In this case the undesirable situation
arises when
$rs<0$ so that we have
$p/r < q/s$, and the chamber ${\bf R}$ lies to the right of the
wall connecting $q/s$ to $p/r$, -- the same side on which the
bound state of the charges given in \refb{eb2} exists.
One can now repeat the argument given for $\wt Q\kdot
\wt P>0$ case to arrive at a diagram similar to that in
Fig.~\ref{f4}, but with the end points of the
walls arranged in the order $q'/s'>q/s>p/r$. The
bound state exists in the region between the walls
connecting $q/s$ to $p/r$ and $q'/s'$ to $p/r$, and not in
the chamber ${\bf R}$.

The case where $\wt Q^2=-2$, $\wt P^2\ge 0$ is related to
the case discussed above by S-duality transformation
$\wt Q\to\wt P$, $\wt P\to -\wt Q$, and so need not be
analyzed separately.

\subsection{$\wt Q^2=\wt P^2=-2$}
\begin{figure}
\leavevmode
\begin{center}
\hbox{ ~ \qquad
\epsfysize=1.5in
\epsfbox{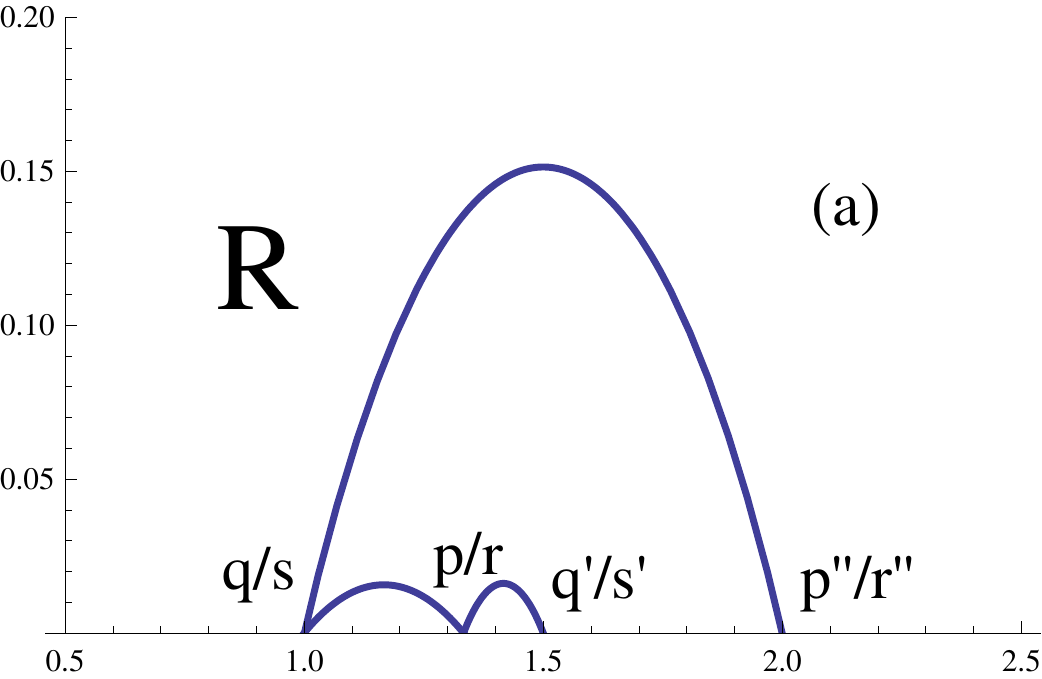} ~ \hfill  ~
\epsfysize=1.5in
\epsfbox{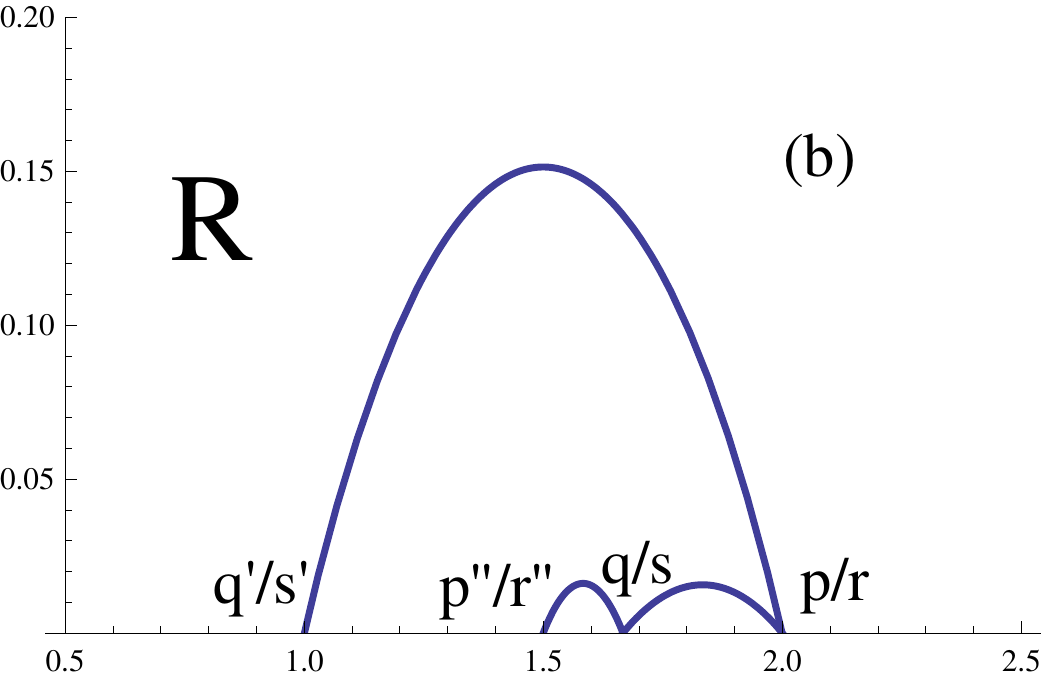}
}
\end{center}
\caption{The arrangement of different walls of marginal
stability for bound states of charges given in \refb{eb2},
\refb{eb25} and \refb{eb26} for $\wt Q^2=\wt P^2=-2$
and $\wt Q\kdot \wt P>0$. Fig.(a) shows the arrangement
for $r>s$ and Fig.(b) shows the arrangement for
$r<s$. In both cases the bound states exist in the region
between the walls and hence not in ${\bf R}$.}
\label{f5}
\end{figure}

First we note that in this case we cannot have $rs=0$,
since if $r=0$ we have $p=s=\pm 1$ and if $s=0$
we have $q=-r = \pm 1$. In both cases eq.\refb{eb3} will
give $P^2=-2$ contradicting the assumption 
that $P^2\le -4$. Thus we choose $rs\ne 0$.
First consider the case
$\wt Q\kdot \wt P>0$. In this case 
for $rs<0$ we have $p/r<q/s$ and the chamber {\bf R} lies
to the right of the wall connecting $q/s$ to $p/r$
as in Fig.~\ref{f3}(a) -- on the opposite side of the domain in
which the bound state exists.
Thus
the problematic case, where the bound state 
exists on the
same side of the wall of marginal stability as the
chamber ${\bf R}$, arises for
\be \label{econd1}
\wt Q\cdot \wt P>0, \qquad rs>0\, ,
\ee
so that
we have $q/s < p/r$ (see Fig.~\ref{f5}(a) and (b)).
We now
consider
two other bound states carrying charges:
\ben \label{eb25}
&& (p' \wt Q', r'\wt Q'), \quad (q' \wt P', s' \wt P')\, ,
\qquad \wt Q' = s'Q - q'P, \quad \wt P' = -r'Q + p' P, 
\nonumber \\ \cr
&& \pmatrix{p' & q'\cr r' & s'} =  \pmatrix{p & q-pu\cr
r & s - ru}, \quad \pmatrix{\wt Q'\cr \wt P'} = \pmatrix{\wt Q+  u
\wt P\cr \wt P}, \quad u\equiv \wt Q\kdot \wt P.
\een
and
\ben \label{eb26}
&& (p'' \wt Q'', r''\wt Q''), \quad (q''\wt P'', s'' \wt P'')\, ,
\qquad \wt Q'' = s''Q - q''P, \quad \wt P'' = -r''Q + p'' P, 
\nonumber \\ \cr
&& \pmatrix{p'' & q''\cr r'' & s''} =  \pmatrix{p-qu & q\cr
r-su & s}, \quad \pmatrix{\wt Q''\cr \wt P''} = 
\pmatrix{\wt Q\cr \wt P+u\wt Q}, \quad 
u\equiv \wt Q\kdot \wt P.
\een
It can be checked that $\wt Q^{\prime 2} = \wt Q^{\prime 
\prime 2}=\wt Q^2$, $\wt P^{\prime 2} = \wt P^{\prime 
\prime 2}=\wt P^2$ and
$\wt Q'\kdot \wt P' = \wt Q'' \kdot \wt P'' = - \wt Q\kdot
\wt P$. Thus these bound states carry the same index
as the bound state of the charges given in \refb{eb2}.
By applying the transformations \refb{eb25} and
\refb{eb26} alternatively (with $u \to -u$ at every step
to account for the change in the sign of $\wt Q\cdot
\wt P$)  we can generate a series of
other bound states with the same index. 
According to the proposal of \S\ref{sbound}
all these bound states represent the same physical 
state.

Since $\wt Q'\kdot \wt P'$ and
$\wt Q''\kdot \wt P''$ are both negative,
the  bound state described in \refb{eb25}
exists to the
right of the wall connecting $q'/s'$ to $p/r$ and the
 bound state given in \refb{eb26}
exists on the right of the wall
connecting $q/s$ to $p''/r''$.  Let us first
consider the case
$r > s$. In this case we must have $u < r/s+1$ since
if $u\ge 1 + r/s$, it follows from \refb{eb3} and
\refb{econd1} that
\be \label{ep2}
P^2 \ge -2r^2 - 2 s^2 + 2 rs \left(1 + {r\over s}\right)
\ge 0\, ,
\ee
contradicting the fact that $P^2\le -4$. Furthermore 
it can be seen as follows that $u$
cannot be equal to $r/s$. Since $u\in \ZZZ$, 
$u=r/s$ will imply that 
$r/s$ must be an integer. Since $r$ and
$s$ are relatively prime this would imply $s=1$, 
$r=u=\wt Q\cdot \wt P$, and
from \refb{eb3} we shall get $P^2 =-2$.
This contradicts the fact that we have chosen $P^2\le -4$.
Thus we are left with two possibilities:
$u< r/s$ and $r/s < u < r/s + 1$.
Let us first assume that $u<r/s$.
Furthermore since $u\ge 1$ and $s/r <1$ we have
$u> s/r$. 
In this case it
follows from \refb{eb25} and
\refb{eb26} that $p''/r'' \ge q'/s'> p/r>q/s$ and
the different points are arranged on the real $\tau$ axis
as shown in Fig.~\ref{f5}(a). 
Since the bound state exists only to the right
of the wall connecting $q/s$ to $p''/r''$, we see that
it does not exist in the chamber ${\bf R}$. 
The case $s>r$ with $u< s/r$ can be
analyzed similarly and leads to the arrangement
described in Fig.~\ref{f5}(b). Since the bound state exists
to the right of the wall connecting $q'/s'$ to $p/r$, we again
see that it does not exist in chamber ${\bf R}$.
This leaves us with the cases $r/s <
s/r < u < s/r +1$ and
$s/r< r/s < u < r/s + 1$.

Before we discuss these cases let us see what happens
when $u=\wt Q\kdot \wt P<0$.
Here the problematic case arises when $rs<0$ so that
we have $q/s > p/r$ and the bound state exists to the right
of the wall connecting $q/s$ to $p/r$ -- the same side
on which the chamber ${\bf R}$ lies. 
Using logic similar to the
one for the $\wt Q\kdot \wt P>0$ case one can argue that
for $|s/r| < |u| < |r/s|$ case the arrangements of
various points are opposite to that given in Fig.~\ref{f5}(a)
and the bound state lies to the
left of the wall connecting $q/s$ to $p''/r''$ so that there
is no bound state in the chamber ${\bf R}$. Similarly for
$|r/s|<|u| < |s/r|$ we can show that 
the arrangements of
various points are opposite to that given in Fig.~\ref{f5}(b)
and the bound state lies to the
left of the wall connecting $q'/s'$ to $p/r$ so that there
is no bound state in the chamber ${\bf R}$.
Thus here again we are left to deal with the cases
$|s/r|< |r/s|<|u|<|r/s|+1$ and 
$|r/s|< |s/r|<|u|<|s/r|+1$, which, taking into
account the negative signs of $u$ and $sr$, 
can be expressed as
$r/s - 1 < u < r/s < s/r$ and $s/r-1<u<s/r<r/s$.

Thus it remains to analyze 
the cases $r/s< s/r < u < s/r +1$ and
$s/r< r/s < u < r/s + 1$ for $sr>0$, $u>0$ and 
$r/s - 1 < u < r/s<s/r$ and $s/r-1<u<s/r<r/s$ 
for $sr<0$, $u<0$.
It can be seen that
\ben \label{eu1}
sr > 0, \quad u>0, 
\quad s/r< 1, \quad  
r/s < u < r/s + 1 &\Rightarrow &r''s''<0, \quad
|r''s''| < rs\cr
sr>0, \quad u>0, \quad s/r>1, \quad 
s/r < u < s/r +1 &\Rightarrow&
r's' <0, \quad |r's'| < rs \cr
sr<0, \quad u\le 0, \quad |s/r|<1, \quad  r/s - 1 < u < r/s
&\Rightarrow& r''s'' > 0, \quad r''s'' < |rs|\cr
sr<0, \quad u\le 0, \quad  |s/r|>1, \quad s/r-1<u<s/r
&\Rightarrow&
r's' >0, \quad r's' < |rs|\, . 
\nonumber \\
\een
The common feature of all these transformations is that
the value of $|rs|$ is reduced under this transformation. 
Given the new values of $(p,q,r,s)$ there are two
possibilities: either $|u|$ lies between  $|r/s|$ and $|r/s|+1$
for $|r|>|s|$ and between $|s/r|$ and $|s/r|+1$ for $|s|>|r|$
or it lies outside this range. In the former case we can apply
\refb{eu1} again to further reduce the value of $|rs|$.
In the second case we can use the results of our
previous analysis to conclude that at the next step
either the wall connecting $q'/s'$ to $p/r$ or the wall
connecting $q/s$ to $p''/r''$ will shield the chamber ${\bf R}$
from the region where the bound state exists. This process
will have to stop eventually since $|rs|$ has a lower
bound of 1, and for $|rs|=1$ we have $|r/s|=|s/r|=1$,
making it impossible to have $|r/s|<|u|<|r/s|+1$ or
$|s/r| < |u| < |s/r|+1$ with non-zero integer $u$.
Thus we shall eventually produce a wall which will
shield from ${\bf R}$ the region in which the bound 
state exists.

This establishes that in the
standard form $P^2\le -4$, there are no two centered
bound states contributing to the index in the chamber ${\bf R}$.
Since we have already seen that the microscopic
index vanishes in this case, we conclude that the
contribution to the index from single centered black holes
must vanish in this case. This is in agreement with the
macroscopic result that there are no single centered black
holes for charge vectors with negative discriminant.

\subsection{Explicit results}

The above discussion has been somewhat abstract, but
it allows us to identify precisely which two centered black
holes contribute to the index in any given chamber. 
For this we 
make an appropriate duality transformation
to bring the charge vector to the `standard form' 
$P^2\le -4$. This takes the original chamber to some
other chamber ${\bf C}$. We can then find an appropriate
path from ${\bf R}$ to ${\bf C}$, 
examine the walls crossed by
this path and compute the contribution 
given in \refb{ew5a} from the 2-centered
black hole solutions which appear as we move from the
chamber ${\bf R}$ 
to ${\bf C}$. The net contribution to the index is
obtained by adding up these contributions. To determine the
constituents in the original duality frame we need to
make an inverse duality transformation to go back to the
original frame. 
We have carried out this analysis for all the
negative discriminant charge vectors 
appearing in table
\ref{t1}. The results are shown in table \ref{t2}.
We can see that the last column agrees with the entries
for the index appearing in table \ref{t1}, confirming that
all the contributions to the index of negative discriminant states
come from two centered black hole solutions.

\sectiono{Generalization to CHL models} \label{sgen}

In this section we shall discuss generalization of
our analysis of \S\ref{sneg} to CHL models.
CHL models are obtained by taking a $\ZZZ_N$ orbifold
of heterotic string theory on $T^6$, where the $Z_N$ acts
as a translation by $2\pi/N$ along one of the circles, and
a $\ZZZ_N$ rotation on the left-movers satisfying the
level matching 
condition\cite{9505054,9506048,9508144,9508154}. 
The index for a class of quarter BPS dyons in these
theories is known 
exactly\cite{0510147,0602254,0603066,0605210,0609109,
0612011} and has a form similar to the one given in
\refb{es3}, \refb{es4} with $\Phi_{10}$ replaced by another
modular form $\Phi$. 
For simplicity we shall restrict our analysis
to the models with prime values of $N$: $N=2,3,5$ and 7.
In this case the main differences in the analysis 
arise due to the following reasons:
\begin{enumerate}
\item The new function  $\Phi$ has period 1 in $\rho$ and
$v$ but period $N$ in $\sigma$. As a result $Q^2$ is
quantized in units of $1/N$.
\item The prefactor $e^{2\pi i (\rho+\sigma+v)}$ is
replaced by $e^{2\pi i(\rho + \sigma/N + v)}$. As a result
in the chamber ${\bf R}$ the lowest values of $P^2$ and
$Q^2$ for which $d(Q,P)$ is non-zero are $-2$ and $-2/N$
respectively. The chamber ${\bf R}$ is defined as before as
the chamber to the right of the wall connecting
0 to $i\infty$ in the $\tau$ plane. In the ($v_2/\sigma_2$,
$\rho_2/\sigma_2$) plane it represents the chamber to the
left of the $v_2=0$ line.
\item The $SL(2,\ZZZ)$ S-duality group appearing
in \refb{ew2} is reduced to $\Gamma_1(N)$
described by the matrices
\be \label{exy1}
\pmatrix{a & b\cr c & d}, \quad ad-bc=1, \quad
\hbox{$c=0$ mod $N$, \, $a,d=1$ mod $N$, 
\, $b\in\ZZZ$}\, .
\ee
In fact it was shown in \cite{0510147} that the modular form
$\Phi$ is actually invariant under the $\Gamma_0(N)$
subgroup of $SL(2,\ZZZ)$ where we relax the conditions
$a,d=1$ mod $N$ to $a,d\in\ZZZ$.
\item The possible 2-centered bound states contributing
to the index arise from configurations described at the end
of \S\ref{scount} with the additional restriction that 
$r\in N\ZZZ$\cite{0702141}. This implies that
\be \label{eimply}
\pmatrix{p & q\cr r & s} \in \Gamma_0(N)\, .
\ee
Furthermore eq.\refb{ew5a} for the index carried by the
bound state is modified to
a different formula 
\be \label{ew5b}
(-1)^{Q\kdot P+1} \, |(sQ-qP).(-rQ+pP)| 
\, f_1((sQ-qP)^2/2) \, f_2((-rQ+pP)^2/2)\, ,
\ee
where $f_1(n)$ and $f_2(n)$ are defined via the equations:
\ben \label{edeff1f2}
q^{-1/N} \prod_{k=1}^\infty (1-q^k)^{-24/(N+1)}
(1-q^{k/N})^{-24/(N+1)}
&=& \sum_{n=-1/N}^\infty f_1(n) q^n\, , \cr
q^{-1} \prod_{k=1}^\infty (1-q^k)^{-24/(N+1)}
(1-q^{kN})^{-24/(N+1)}
&=& \sum_{n=-1}^\infty f_2(n) q^n
\, .
\een
For $N=1$, $f_1$ and $f_2$ both reduce to $f(n)$
defined in eq.\refb{ew4} and we recover the result
for heterotic string theory on $T^6$.
\end{enumerate}

Let us now reexamine our analysis
of \S\ref{sneg} for these models. First of all we need
to show that we can bring a charge vector $(Q,P)$ to the
standard form where $P^2\le -4$ so that the index of the
state vanishes in the chamber ${\bf R}$. 
If $Q^2<0$, then by making a
$P\to P + K N Q$, $Q\to Q$ transformation 
we get $P^2\to P^2 + 2KN Q\kdot P
+ K^2 N^2 Q^2$, and this can be made arbitrarily large
negative for sufficiently large integer
$K$. For $Q^2=0$ the same
transformation works if we take $K$ to have the opposite sign
of $Q\cdot P$. For $Q^2>0$ we 
consider a $\Gamma_1(N)$
matrix with the following values of $c$, $d$:
\be \label{exy2}
c = -Q\kdot P N L K, \quad d = Q^2 N L K + 1\, ,
\qquad L = \hbox{l.c.m}\, (Q\kdot P, Q^2)\, ,
\ee
where $K$ is a large integer and l.c.m. stands for the
lowest common multiple. In this case 
$c$ and $d$
cannot contain a common factor since all prime factors
of $c$ are also prime factors of $(d-1)$ by construction.
Thus there exist $a$ and $b$ such that 
$\pmatrix{a & b\cr
c & d}\in SL(2,\ZZZ)$. Furthermore by construction
$c=0$ mod $N$ and $d=1$ mod $N$, and hence $ad-bc=1$
gives $a=1$ mod $N$. Thus $\pmatrix{a & b\cr c & d}
\in \Gamma_1(N)$ and is
an allowed duality transformation. It now follows from 
\refb{ew2} and \refb{exy2} that under this transformation
\be \label{exy3}
P^2 \to K^2 \left[
L^2 N^2 Q^2 \left(Q^2 P^2 - (Q\kdot P)^2\right)
+ \OO(K^{-1})\right]\, .
\ee 
Thus by taking $K$ to be sufficiently large and using the
fact that $ Q^2 P^2 - (Q\kdot P)^2<0$ we can make
$P^2$ arbitrarily large and negative. This allows us to
bring $(Q,P)$ to the standard form $P^2\le -4$.

Thus we now need to show following the analysis of
\S\ref{sneg} that for the new charge vector in the
chamber ${\bf R}$ there are no two centered bound states.
We consider the possible bound states of the form
\refb{eb2} with $\pmatrix{p & q\cr r & s}\in \Gamma_0(N)$.
The case $\wt P^2\ge 0$, $\wt Q^2\ge 0$ proceeds as
in the $T^6$ case. The case $\wt P^2=-2$, $\wt Q^2\ge 0$
is also identical, -- the only point to note is that the 
transformation $\wt Q' = \wt Q+u\wt P$, $\wt P'=\wt P$
with $u=Q\kdot P$ correspond to multiplcation by the matrix
$\pmatrix{1 & u\cr 0 & 1}\in \Gamma_1(N)$ and hence is
an allowed duality transformation. The cases 
$\wt Q^2=-2/N$ needs a different analysis however.
In this case a transformation $\wt Q \to \wt Q$, $\wt P\to
\wt P + u \wt Q$ appearing in \refb{eb26}
is neither a valid duality transformation,
nor does it leave $\wt P^2$ invariant. Instead we use the
transformation is $\wt Q\to \wt Q$, $\wt P\to \wt P
+ N u \wt Q$. This is a valid duality transformation
and has the effect of leaving $\wt Q^2$
and $\wt P^2$ invariant, and changing the sign of $\wt Q
\cdot \wt P$. Thus we identify bound states 
of constituents related by this transformation.
With this hypothesis we find, after a somewhat lengthy
analysis along the lines of \S\ref{sneg}, 
that there are no two centered bound states
in the chamber ${\bf R}$ for $P^2\le -4$.
This is turn establishes that the negative discriminant
states in any chamber of the moduli space can be accounted
for by the contribution to the index from multi-centered
black holes.

\sectiono{Conclusion} \label{sconc}

Our analysis shows that
the negative discriminant states
in the microscopic spectrum of  a class of
$\NN=4$ supersymmetric string
theories can be accounted for as bound states of 
two centered black holes. This is consistent with the fact
that single centered black hole solutions 
exist only for charges
with positive
discriminant and shows that the description of the system
as a black hole can capture exact information on the
system even for finite charges. This in turn suggests that
quantum gravity in the near horizon geometry may provide
an exact dual description of the system instead of just being
an emergent description that works only for large charges.
The result of \cite{1008.4209} showing that 
quantum gravity
in the the near horizon  geometry
correctly predicts the sign of the index also points
to the same conclusion.

It will clearly be interesting to extend this analysis to
states carrying torsion ($I$ defined in eq.\refb{es2})  
larger
than 1. The bound state metamorphosis rules described
in \S\ref{sbound}, \S\ref{sneg} are likely to be more
complicated for these states. It will also be useful to
understand the origin of
these rules directly from the classical analysis
of two centered solutions.

Another direction that should be explored is 
the macroscopic
origin of the states with zero discriminant. There are
plenty of examples ({\it e.g.} all states carrying 
$Q^2=P^2=Q\cdot P$ in table \ref{t1}) 
for which the index does not
vanish in the chamber {\bf R}, but there are no two
centered configurations contributing to this index.
Smooth horizonless classical solutions constructed
along the lines of \cite{0701216} might play a role
in providing the macroscopic description of these
states.

{\bf Acknowledgement: } 
This work was supported in part by the
J. C.  Bose fellowship of the Department of
Science and Technology, India and the
project 11-R\& D-HRI-5.02-0304.

\end{document}